\documentclass{aa}
\usepackage{graphicx}

\begin{document}
\title{Detached white dwarf main-sequence star binaries}
\author{B. Willems\thanks{Present address: Northwestern University,
    Department of Physics and Astronomy, 2145 Sheridan Road, Evanston,
  IL 60208, USA} \and U. Kolb} 
\offprints{B.\ Willems}
\institute{Department of Physics and Astronomy, The Open University,
Walton Hall, Milton Keynes, MK7 6AA, UK \\
\email{b-willems@northwestern.edu, U.C.Kolb@open.ac.uk} }
\date{Received date; accepted date}

\abstract{We initiated a comprehensive state of the art binary
  population synthesis study of white dwarf main-sequence star (WDMS)
  binaries to serve as a foundation for subsequent studies on
  pre-cataclysmic variables, double white dwarfs, and white dwarf +
  B-star binaries. We considered seven distinct formation channels
  subdivided into three main groups according to the evolutionary
  process that gives rise to the formation of the white dwarf or its
  helium-star progenitor: dynamically stable Roche-lobe overflow
  (Algol-type evolution), dynamically unstable Roche-lobe overflow
  (common-envelope evolution), or stellar winds (single star
  evolution). For each formation channel, we examine the sensitivity
  of the population to changes in the amount of mass lost from the
  system during dynamically stable Roche-lobe overflow, the
  common-envelope ejection efficiency, and the initial mass ratio or
  initial secondary mass distribution. In the case of a flat initial
  mass ratio distribution, the local space density of WDMS binaries is
  of the order of $\sim 10^{-3}\, {\rm pc}^{-3}$. This number
  decreases to $\sim 10^{-4}\, {\rm pc}^{-3}$ when the initial
  mass ratio distribution is approximately proportional to the inverse
  of the initial mass ratio. More than 75\% of the WDMS binary
  population originates from wide systems in which both components
  essentially evolve as if they were single stars. The remaining part
  of the population is dominated by systems in which the white dwarf
  is formed in a common-envelope phase when the primary ascends the
  first giant branch or the asymptotic giant branch. When dynamically
  stable mass transfer proceeds highly conservative and the
  common-envelope ejection process is very efficient, the birthrate of
  WDMS binaries forming through a common-envelope phase is about 10
  times larger than the birthrate of WDMS binaries forming through a
  stable Roche-lobe overflow phase. The ratio of the number of helium
  white dwarf systems to the number of carbon/oxygen or
  oxygen/neon/magnesium white dwarf systems derived from large samples
  of observed WDMS binaries by, e.g., future planet-search missions
  such as SuperWASP, COROT, and Kepler may furthermore constrain the
  common-envelope ejection efficiency.
}

\maketitle

\section{Introduction}

In recent years, substantial numbers of detached white dwarf
main-sequence star (WDMS) binaries have been detected in large-scale
surveys searching for pre-cataclysmic variables (Hillwig et al. 2000,
Raymond et al. 2003), double degenerates (Saffer et al. 1998),
low-mass white dwarfs (Marsh et al. 1995, Maxted et al. 2000),
planetary nebulae (Bond \& Livio 1990, Livio 1992), or even the dark
matter content of the Galaxy's heavy halo (Silvestri et
al. 2002). Upcoming surveys looking for transiting extrasolar planets
are furthermore expected to contribute further to this rapidly growing
database (Farmer \& Agol 2003). So far, the majority of WDMS binaries
found consist of a white dwarf with a low-mass companion. White dwarfs
in binaries with main-sequence secondaries more massive than
$0.3\,M_\odot$ are generally harder to detect due to the large
luminosity difference between the component stars (Iben et al. 1997,
Marsh 2000). Despite this obstacle, several claims have recently been
made on the possible detection of white dwarf binaries with bright
companions of spectral types as early as B(e) (e.g. Holberg et
al. 1998, Vennes et al. 1998, Burleigh \& Barstow 2000, Burleigh et
al. 2001, Shobbrook et al. 2003; and references therein).

WDMS binaries with early B(e)-type companions more massive than $\sim
10\,M_\odot$ are thought to be the end product of binaries in which a
giant-type star transfers mass to its initially less massive
companion. The mass-transfer phase is responsible for exposing the
giant's core as a white dwarf and for increasing the mass of the
secondary so that it evolves towards earlier spectral types. In
addition, transport of angular momentum may spin the secondary up to
form a rapidly rotating Be star (Waters et al. 1988, Pols et
al. 1991). The predicted number of WDMS binaries forming in this way
is, however, rather small (e.g. de Kool \& Ritter 1993, Figs. 2 and 4)
which may pose a problem if the recent claims on the detection of
white dwarf + B-star binaries are confirmed. In addition, as mass
transfer is expected to increase the orbital period it is hard to
explain the formation of short-period white dwarf + early B-star
candidates such as HD\,161103 ($P_{\rm orb} = 4.7$ days, Shobbrook et 
al. 2003) through this evolutionary channel.

The wealth of existing and expected new data combined with new
up-to-date stellar evolution tools (see Hurley et al. 2000) render the 
WDMS binaries a potentially fruitful subject of a comprehensive 
binary population synthesis study. A systematic exploration of the
formation channels in particular may shed new insights on the origin
of short-period white dwarf + B-star binaries. Other interesting
subclasses of detached WDMS binaries include pre-cataclysmic variables
and progenitors of double degenerates.

To the best of our knowledge, de Kool \& Ritter (1993) were the first
to devote an entire study to the population of WDMS binaries as a
class on its own. The authors used a Monte Carlo type simulation in
which the initial binary parameters were randomly generated from
observed orbital period, primary mass, and secondary mass
distributions; and the evolution of the binaries was approximated by
analytical fits to detailed stellar evolution tracks as described by
Politano (1988) and de Kool (1992). The method resulted in
distribution functions for the formation rates and space densities of
detached WDMS binaries (or their direct progenitors) as a function of
their expected orbital and stellar parameters.

In this paper, our aim is to update the population synthesis study by
de Kool \& Ritter (1993) by using the more recent and much more
detailed analytical fits to stellar evolution derived by Hurley et
al. (2000). We study in more detail the differences between the
evolutionary channels leading to the formation of WDMS binaries, and
derive the formation rates and absolute numbers of systems for each of
the formation channels separately. In addition, we consider a much
more elaborate set of population synthesis models characterised by
different assumptions about the fate of the mass transferred during
dynamically stable Roche-lobe overflow, different common-envelope
ejection efficiencies, and different initial mass ratio or secondary
mass distributions.

Our method furthermore differs from that by de Kool \& Ritter (1993)
in the implementation of the observed distributions of the initial
masses and orbital periods: instead of randomly generating them from
the observed distribution functions at the onset of the binary
evolution calculations, we start our simulations from a logarithmically
spaced 3-dimensional grid of initial masses and orbital periods in
which each set of binary parameters has an equal probability. The
likelihood of the different initial binary parameters is then taken
into account by weighting the contribution of each WDMS binary found
in the simulations according to the adopted primary mass distribution,
initial mass ratio or secondary mass distribution, and initial orbital
period distribution.  In doing so, we obtain a better and more uniform
sampling of the entire parameter space, which particularly benefits
the resolution of low-density tails in the distribution functions
describing the masses and orbital periods of WDMS binary
components. This procedure also allows us to vary the
distribution functions for the initial masses and orbital periods
without having to repeat the binary evolution calculations. 

The results of this study are intended to serve as a foundation for more
focused population synthesis studies of white dwarf + B-star binaries
(Willems et al., in preparation), double white dwarfs (Willems et
al., in preparation), and pre-cataclysmic variables. Hence, we here 
concentrate on the effect and the uncertainties associated with the
different model parameters in general. In subsequent papers we hope to
constrain these uncertainties by comparing our results with different
types of WDMS binaries and WDMS binary descendants. The results
presented here therefore do not necessarily represent the best
possible fit to observationally derived binary parameters.

The plan of the paper is as follows. In Section~2, we present the
basic ingredients and the assumptions adopted in the binary population
synthesis calculations. In Section~3, we describe the different
evolutionary channels leading to the formation of WDMS binaries as
well as the ranges of stellar masses and orbital periods available to
each formation channel. Section~4 deals with the effects of
varying degrees of non-conservative mass transfer and different
common-envelope ejection efficiencies on the population of WDMS
binaries. In Section~5, we estimate the order of magnitude of the
formation rates and the expected number of WDMS binaries currently
populating the Galaxy. Section~6 briefly addresses the expected
luminosity differences between the white dwarf and its companion, as
well as their expected orbital radial-velocity variations. A brief 
summary of our results and some concluding remarks are presented in 
Section~7.

\section{Basic concepts and assumptions}
\label{basic}

\subsection{The population synthesis code}

We use the BiSEPS binary population synthesis code introduced by
Willems \& Kolb (2002) to study the formation of detached white dwarf
main-sequence star binaries. The code uses the single star evolution
formulae derived by Hurley et al. (2000) and follows the main
steps of the binary evolution scheme outlined by Hurley et al. 
(2002). All binary orbits are treated as circular and stellar rotation
rates are kept synchronised with the orbital motion at all times. For
the purpose of this investigation, we furthermore limit ourselves to
Population~I stellar compositions.

When a binary component overflows its Roche-lobe, the stability of the
ensuing mass-transfer phase is determined by means of the radius-mass
exponents introduced by Webbink (1985). Mass transfer taking place on
the dynamical time scale of the donor star is assumed to lead to a
common-envelope (CE) phase during which the orbital separation is
reduced and the envelope is expelled from the system. The phase is
modelled in the usual way by equating the binding energy of the donor
star's envelope to the change in the orbital energy of the binary
components. The orbital separation $a_{\rm f}$ at the end of the
common-envelope phase is then related to the orbital separation
$a_{\rm i}$ at the beginning of the phase as
\begin{equation}
{{a_{\rm f}} \over {a_{\rm i}}} = {{M_{\rm c}/M_1} \over
  {1 + 2\, M_{\rm e}/\left( M_2\, \alpha_{\rm CE}\, \lambda\,
  r_{\rm L} \right)}}. \label{ce}
\end{equation}
Here $G$ is the gravitational constant, $M_{\rm c}$ and $M_{\rm e}$
are the core and envelope mass of the Roche-lobe overflowing star,
$M_1 = M_{\rm c} + M_{\rm e}$ is the total mass of the donor star,
$M_2$ is the mass of the companion, and $r_{\rm L}=R_{\rm L}/a_{\rm
i}$ is the radius of the donor star's Roche lobe in units of the
orbital separation at the start of the common-envelope phase. For the
binding-energy parameter $\lambda$ and the common-envelope ejection
efficiency $\alpha_{\rm CE}$, we adopt the commonly used values
$\lambda=0.5$ and $\alpha_{\rm CE}=1.0$ (e.g.\ de Kool 1992, Politano
1996).

In the case of dynamically stable mass transfer, a fraction
$1-\gamma_{\rm RLOF}$ of the transferred mass is assumed to be
accreted by the donor star's companion, while the remaining fraction
$\gamma_{\rm RLOF}$ is lost from the system carrying away the specific
orbital angular momentum of the companion. The amount of angular
momentum carried away by matter leaving the system during
non-conservative Roche-lobe overflow is, however, still an unresolved
issue, so that this quantity is effectively a free parameter (see,
e.g., the appendix in Kolb et al. 2001).

For non-degenerate accretors, we set
\begin{equation}
1 - \gamma_{\rm RLOF} = \min \left( 10\, {\tau_{\dot{M}} \over
  \tau_{\rm HK,a}} , 1 \right),  \label{gam}
\end{equation}
where $\tau_{\dot{M}}$ is the mass-transfer time scale of the donor
and $\tau_{\rm HK,a}$ is the thermal time scale of the accretor. With
this prescription, the mass-transfer phase is conservative as long as
$\tau_{\rm HK,a} < 10\, \tau_{\dot{M}}$ (see also, for example, Iben
\& Tutukov 1987, Pols et al. 1991, Hurley et al. 2002). Accretion onto
white dwarfs is assumed to be fully non-conservative so that
$\gamma_{\rm RLOF}=1$ and $\dot{M}_{\rm WD}=0$, where $M_{\rm WD}$ is
the mass of the white dwarf. This assumption does not affect the
formation of detached WDMS binaries, but may influence the fate of
their descendants when the main-sequence star becomes larger than its
critical Roche lobe. We do not deal with neutron star and black hole
accretors in this investigation.

For more details on the treatment of mass-loss and mass-accretion in
the BiSEPS code we refer to Willems \& Kolb (2002).

\subsection{Initial masses and orbital periods}
\label{init}

We start our population synthesis study by evolving a large number of
binaries initially consisting of two zero-age main-sequence stars with
a mass between $0.1$ and $30\,M_\odot$ and an orbital period between
$0.1$ and $100\,000$ days. The initial primary and secondary masses,
$M_1$ and $M_2$, and the initial orbital periods $P_{\rm orb}$ are
taken from a grid consisting of 60 logarithmically spaced stellar
masses and 300 logarithmically spaced orbital periods. The maximum
evolutionary age considered for each binary is 10\,Gyr. For symmetry
reasons only binaries with $M_1 > M_2$ are evolved.

The number of systems following a sequence of evolutionary phases
similar to those of a given binary in our simulated sample is
determined by the probability of the binary's initial parameters, by
the star formation rate at the birth of the zero-age main-sequence
binary, and by the fraction of stars in binaries. We assume the
initial primary masses to be distributed according to the normalised
initial mass function (IMF)\footnote{We note that the adopted IMF is a
simplified version of the IMF by Kroupa et al. (1993). The
simplification is introduced because of the still existing
uncertainties in the IMF for low-mass stars (see, e.g., Scalo 1998,
Kroupa 2001). Its effect on our results is small in comparison to the
overall uncertainties of the population synthesis models.}
\begin{equation}
\renewcommand{\arraystretch}{1.4} 
\xi \left(M_1 \right) = \left\{
  \begin{array}{ll}
  0 & \hspace{0.3cm} M_1/M_\odot < 0.1, \\ 
  0.38415\, M_1^{-1} & \hspace{0.3cm} 0.1 \le M_1/M_\odot < 0.75, \\ 
  0.23556\, M_1^{-2.7} & \hspace{0.3cm} 0.75 \le M_1/M_\odot < \infty,
  \end{array}
\right. \label{imf}
\end{equation}
the initial mass ratios $q=M_2/M_1$ according to
\begin{equation}
\renewcommand{\arraystretch}{1.4} 
n(q) = \left\{
  \begin{array}{ll}
  \mu\,q^\nu & \hspace{0.3cm} 0 < q \le 1, \\ 
  0 & \hspace{0.3cm} q > 1,
  \end{array}
\right. \label{imrd}
\end{equation}
and the initial orbital separations $a$ according to
\begin{equation}
\renewcommand{\arraystretch}{1.4} 
\chi (a) = \left\{
  \begin{array}{ll}
  0 & a/R_\odot < 3 \mbox{ or } a/R_\odot > 10^6, \\ 
  0.078636\, a^{-1} & 3 \le a/R_\odot \le 10^6.
  \end{array}
\right. \label{iosd}
\end{equation}
In Eq.~(\ref{imrd}), $\nu$ is a constant and $\mu$ a normalisation
factor depending on $\nu$. Unless stated otherwise, we set $\nu=0$ and
$\mu=1$. The upper limit of $10^6\,R_\odot$ in the distribution
of the initial orbital separations is larger than in our previous
investigations in order to properly take into account the contribution
of very wide systems to the population of WDMS binaries. For more
details and references on the adopted distribution functions, we refer
to Willems \& Kolb (2002).

We furthermore assume all stars to be in binaries and adopt a constant
star-formation rate $S$ calibrated so that one binary with $M_1 >
0.8\,M_\odot$ is born in the Galaxy each year (see also Iben \&
Tutukov 1984, Han et al. 1995, Hurley et al. 2002). When combined with
an effective Galactic volume of $5 \times 10^{11}\, {\rm pc^3}$, this
yields an average local birthrate of Galactic white dwarfs of $2
\times 10^{-12}\,{\rm pc^{-3}\, yr^{-1}}$, which is consistent with
observations (Weidemann 1990). From this calibration, it follows that 
\begin{equation}
S\, \int_{0.8}^{\infty} \xi \left(M_1 \right) dM_1 = 1, \label{sfr}
\end{equation}
so that $S=4.9$ ${\rm yr^{-1}}$. This rate may be converted into an
approximate local star formation rate (expressed in ${\rm pc^{-3}\,
  yr^{-1}}$) by dividing it by $5 \times 10^{11}\, {\rm pc^3}$.  As
the star formation rate may have been higher in the past, the
calibration of the rate to match the observationally inferred current
birthrate of Galactic white dwarfs may yield an underestimate of the
number of binaries with old component stars (Boissier \& Prantzos
1999).

\section{Formation channels}
\label{form}

Stars in close binaries can evolve into white dwarfs either through
the loss of their envelope by the action of a stellar wind or by mass
transfer resulting from dynamically stable or unstable Roche-lobe
overflow. For brevity, we refer to binaries that do not undergo
mass-transfer episodes as non-interacting binaries, even if some mass
exchange and orbital evolution takes place due to the action of a
stellar wind. Binaries in which a white dwarf or its direct progenitor
is formed as the end product of mass transfer are referred to as
interacting binaries. In what follows, we divide the latter group
according to the stability of the mass-transfer phase and according to
the remnant it leaves behind.

Since detached WDMS binaries constitute an important intermediary
phase in the formation of many more exotic binary systems such as
cataclysmic variables and double degenerates, some of the formation
channels described below have been partially discussed before, albeit
maybe in lesser detail and in different contexts.  The evolution of
binaries with low- to intermediate-mass component stars has been
studied in detail by, e.g., Iben \& Tutukov (1985, 1986a, 1987), van
der Linden (1987), de Loore \& Vanbeveren (1995), Langer et
al. (2000), Han et al. (2000), Nelson \& Eggleton (2001), Chen \& Han
(2002, 2003); and references therein.

\subsection{Dynamically stable mass transfer}
\label{stable}

\subsubsection{Case B RLOF with a white dwarf remnant}
\label{sch1}

The first formation channel applies to initial binaries consisting of
two low-mass main-sequence stars with orbital periods that are too
short to allow the primary to evolve on the giant branch
without overflowing its Roche lobe. Most of the systems initiate mass
transfer on the thermal time scale of the donor star when it
approaches the end of the main sequence or when it crosses the
Hertzsprung gap. Once the mass ratio is inverted, mass transfer
generally slows down and continues as the donor star ascends the first
giant branch. In most cases, the case~B mass-transfer phase is more
important for determining the subsequent evolution than the initial
case~A phase. For brevity, we therefore simply refer to the case~B
phase as the phase characterising the formation channel.  

\begin{figure}
\resizebox{8.4cm}{!}{\includegraphics{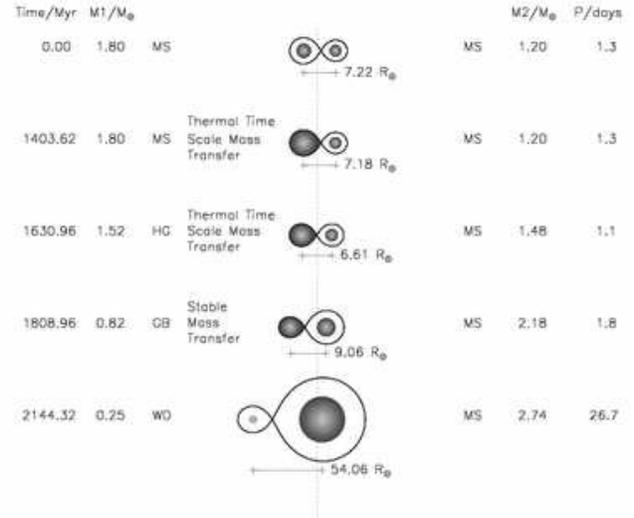}}
\caption{Schematic representation of the main evolutionary phases
  leading to the formation of a WDMS binary via evolutionary channel~1
  (see text for details). MS stands for main sequence, HG for
  Hertzsprung gap, GB for giant branch, and WD for white
  dwarf.}
\label{ch1}
\end{figure}

Since the thermal time scale of the accreting companion
is of the same order of magnitude as that of the donor star, the
mass-transfer phase is highly conservative [see Eq.~(\ref{gam})]. The
secondary's mass and the orbital period may therefore increase
substantially with respect to their values at the onset of the
Roche-lobe overflow phase. The mass-transfer phase terminates when the
giant's entire envelope is transferred to the companion, exposing its
helium core as a low-mass white dwarf. The main evolutionary phases
characterising this formation channel are summarised schematically in
Fig.~\ref{ch1}. For future reference, we label this channel as
channel~1.

\begin{figure*}
\begin{center}
\resizebox{12.6cm}{!}{\includegraphics{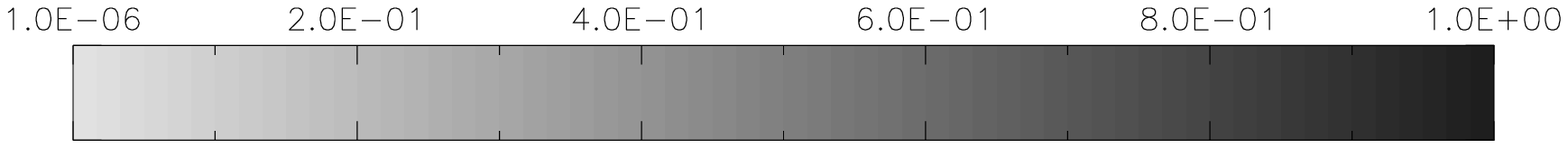}} \\
\resizebox{8.4cm}{!}{\includegraphics{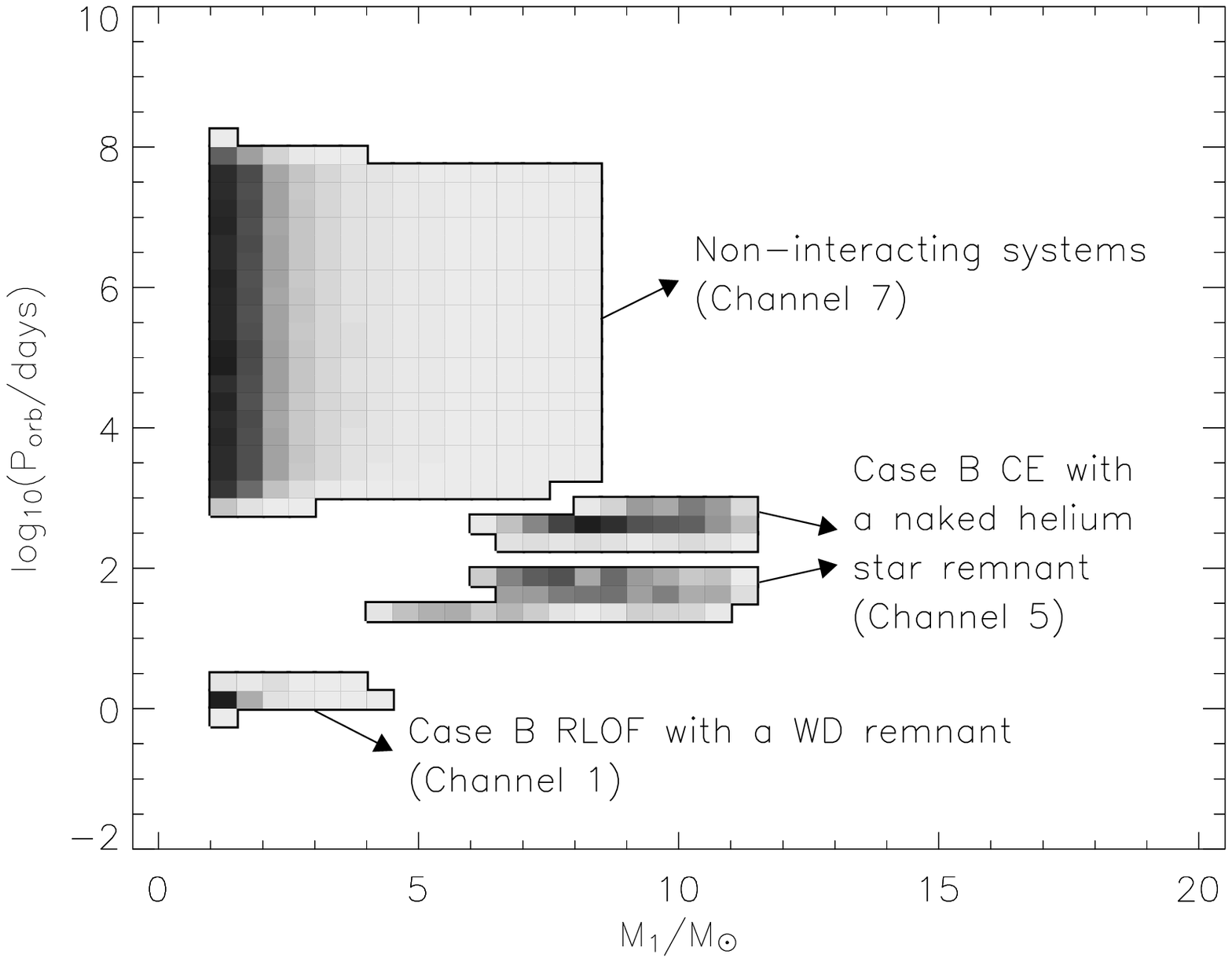}}
\resizebox{8.4cm}{!}{\includegraphics{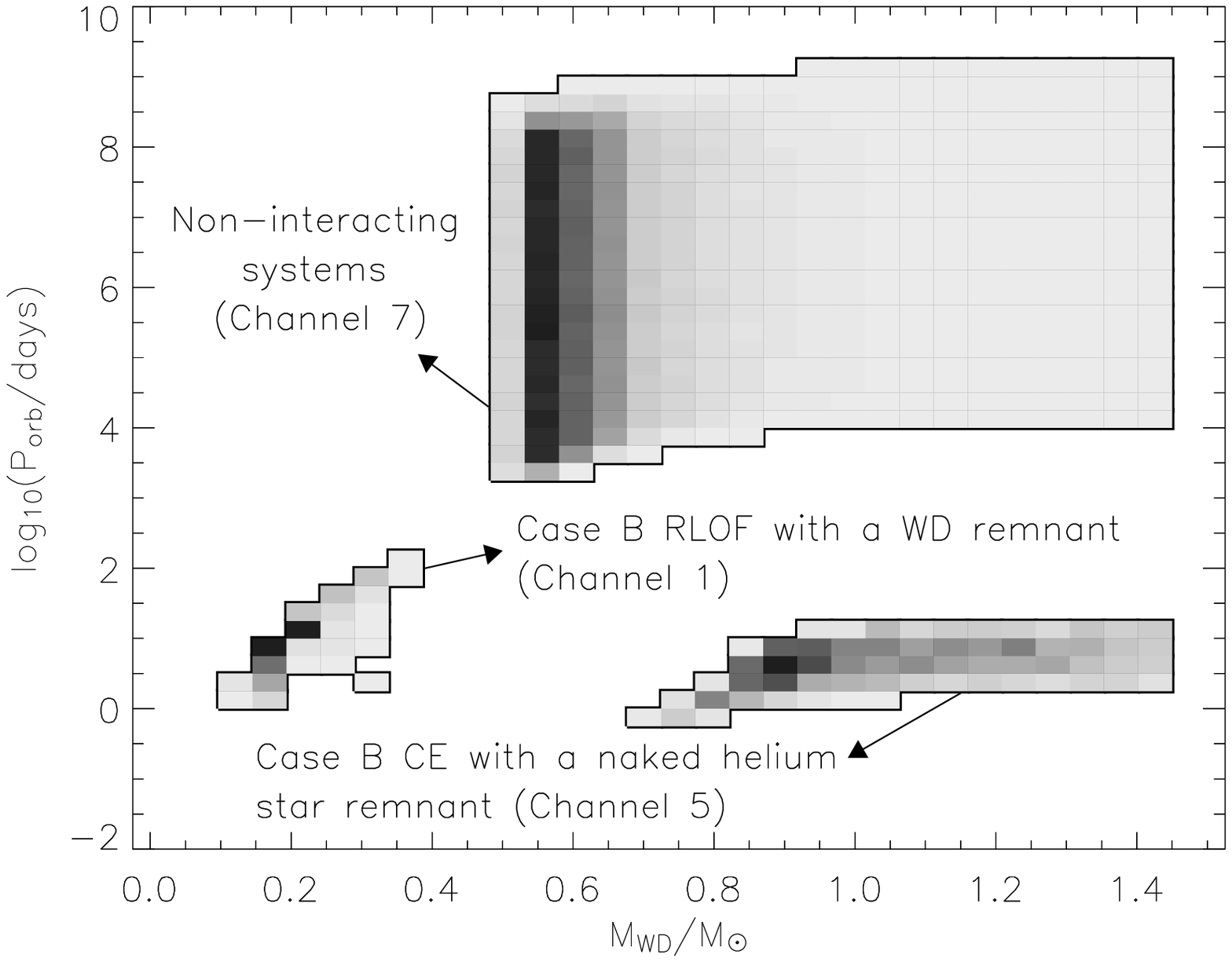}} \\
\resizebox{8.4cm}{!}{\includegraphics{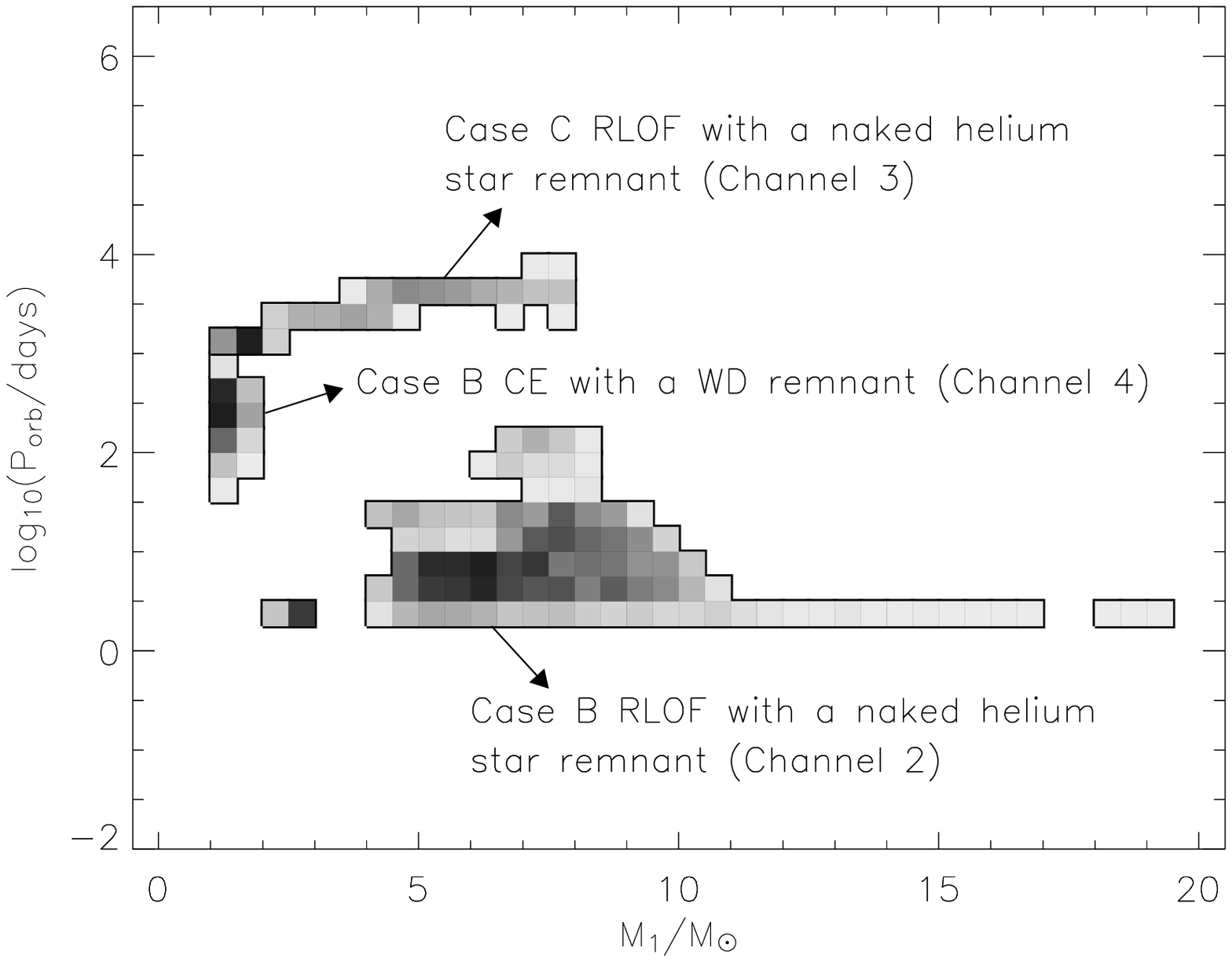}}
\resizebox{8.4cm}{!}{\includegraphics{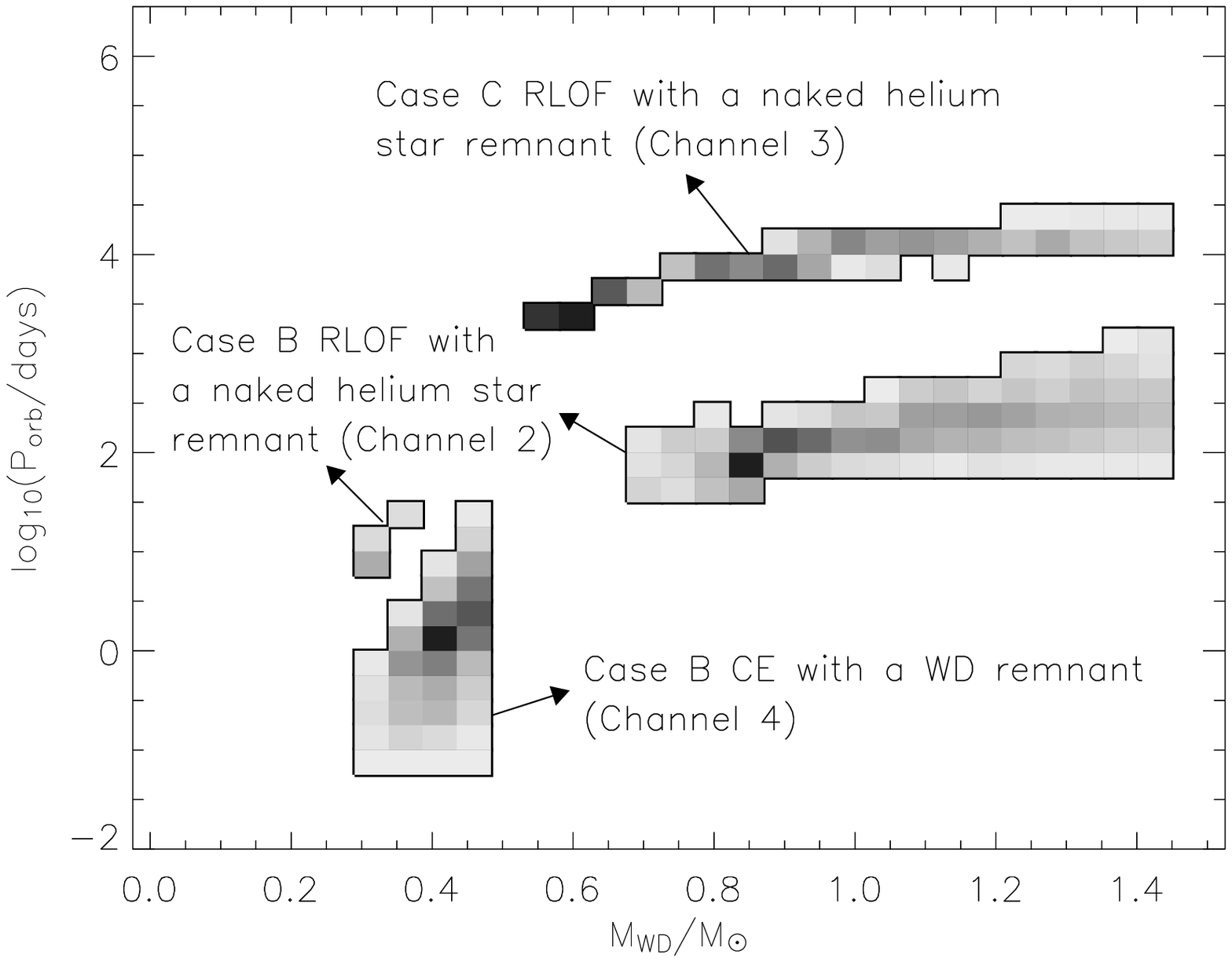}} \\
\resizebox{8.4cm}{!}{\includegraphics{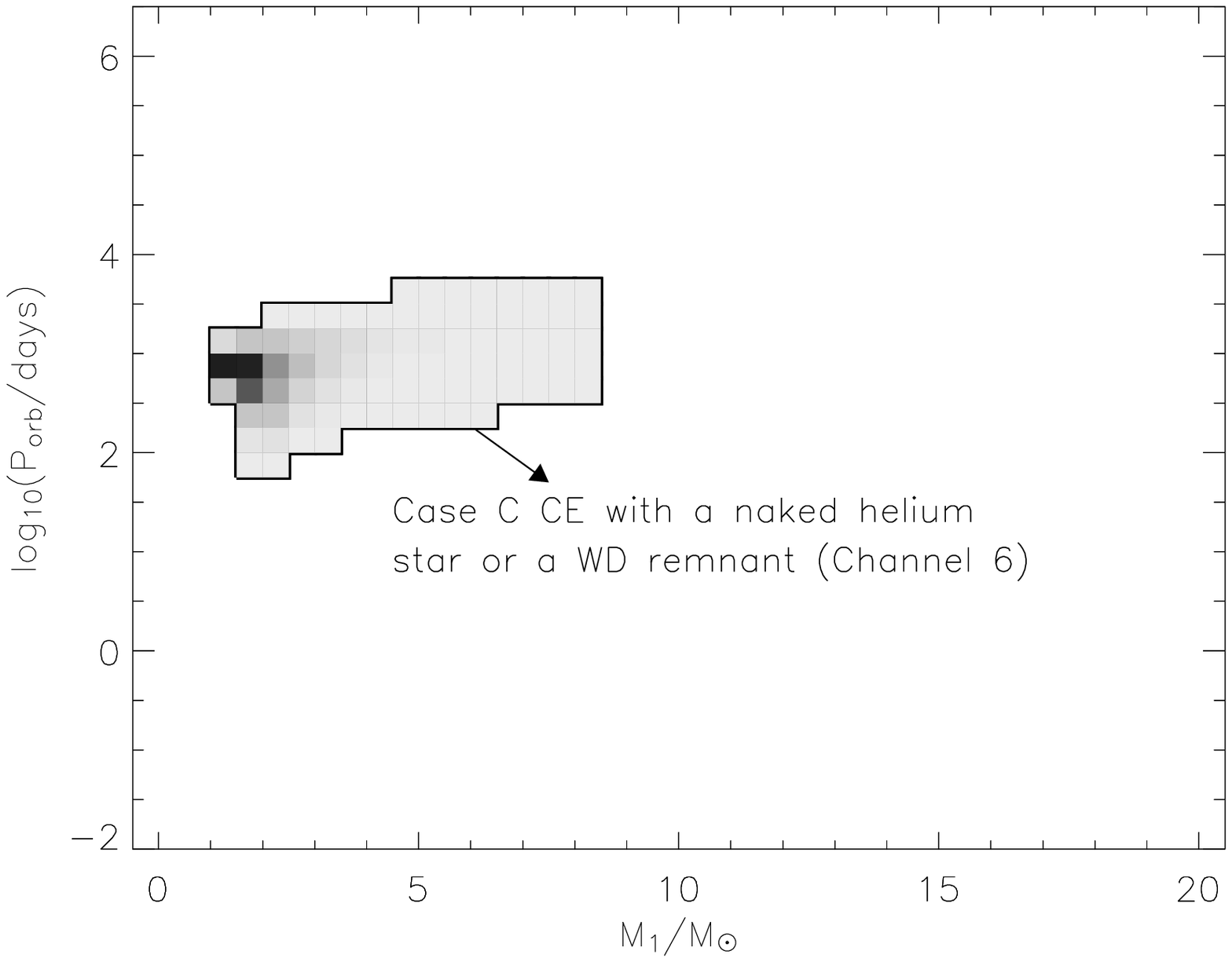}}
\resizebox{8.4cm}{!}{\includegraphics{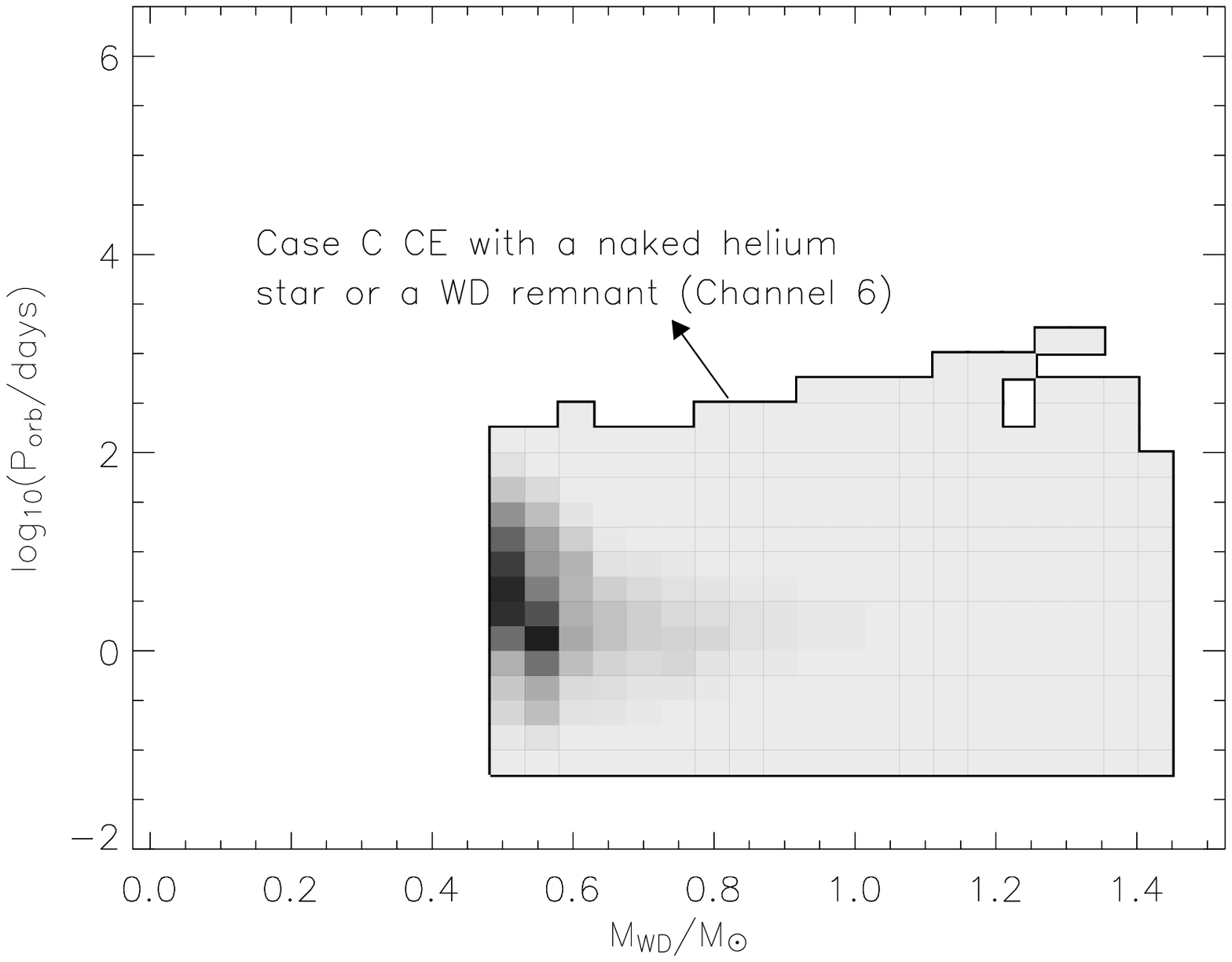}} \\
\caption{Distribution of present-day WDMS binaries in the $\left( M_1,
  P_{\rm orb} \right)$-plane at the beginning of the progenitors'
  evolution (left-hand panels) and at the beginning of the WDMS binary
  phase (right-hand panels). The distributions are normalised so that
  for each formation channel the bin containing the largest
  contribution to the population has a PDF value equal to one.}
\label{2DM1}
\end{center}
\end{figure*}

\begin{figure*}
\begin{center}
\resizebox{12.6cm}{!}{\includegraphics{bwbar0to1.ps}} \\
\resizebox{8.4cm}{!}{\includegraphics{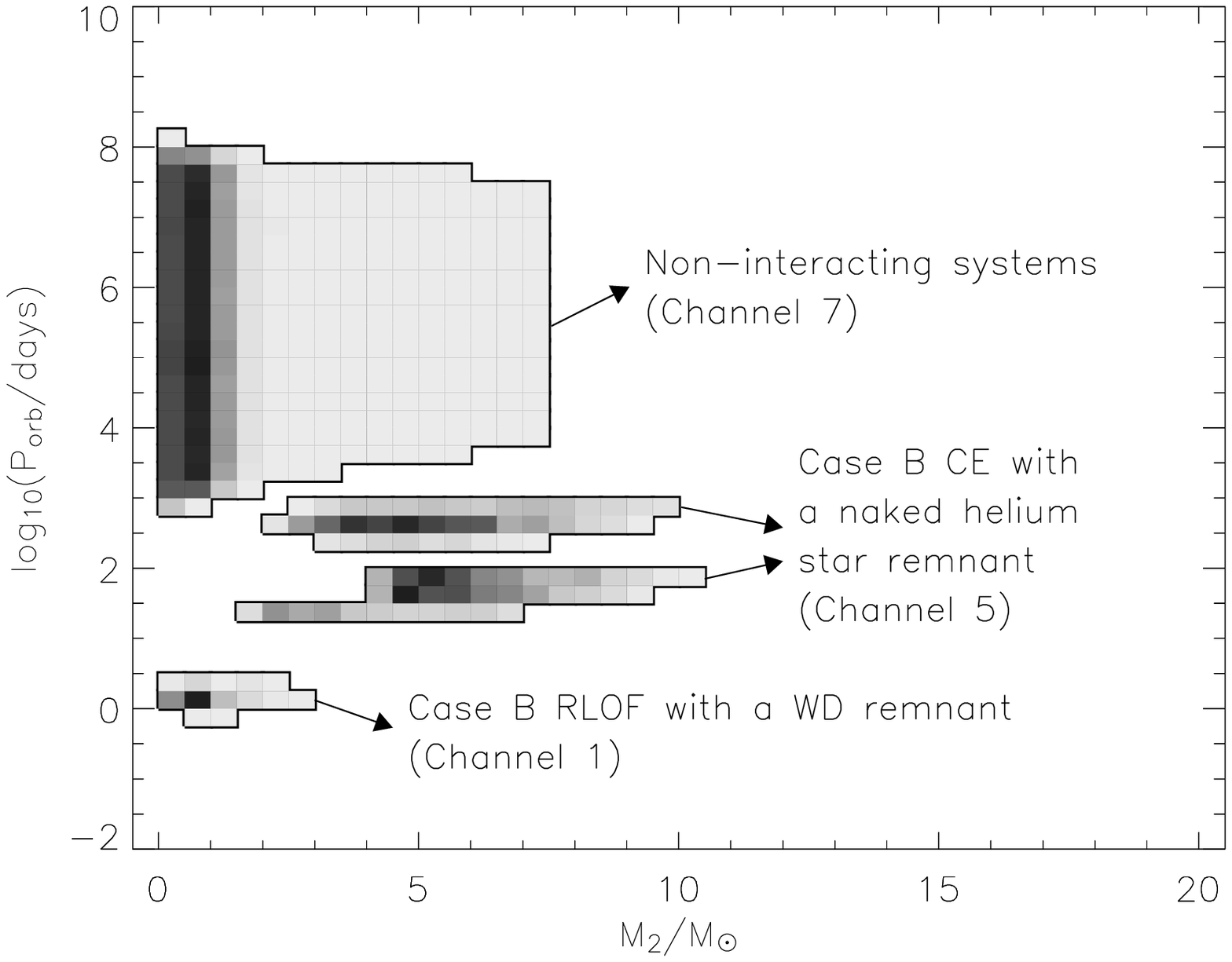}}
\resizebox{8.4cm}{!}{\includegraphics{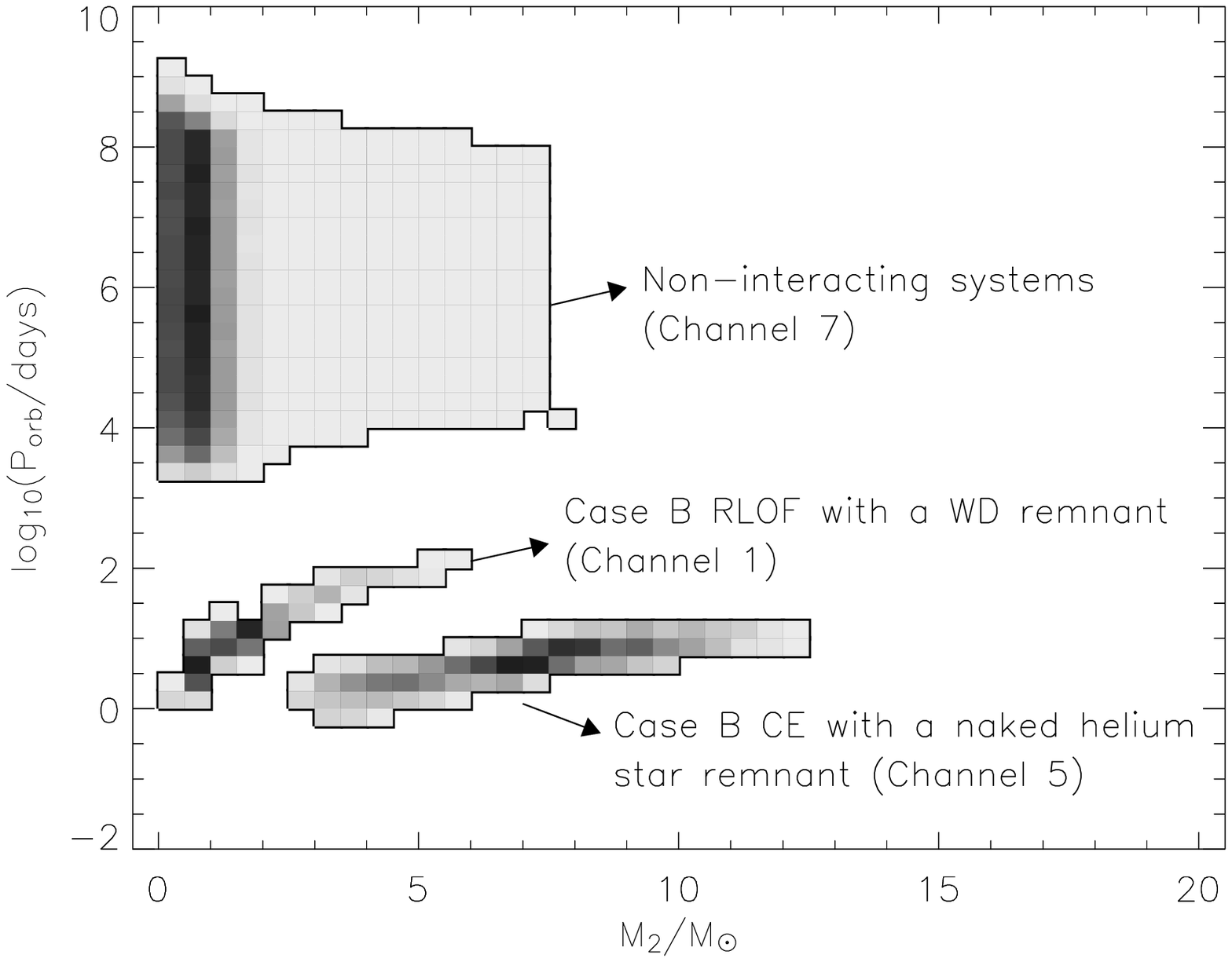}}
\resizebox{8.4cm}{!}{\includegraphics{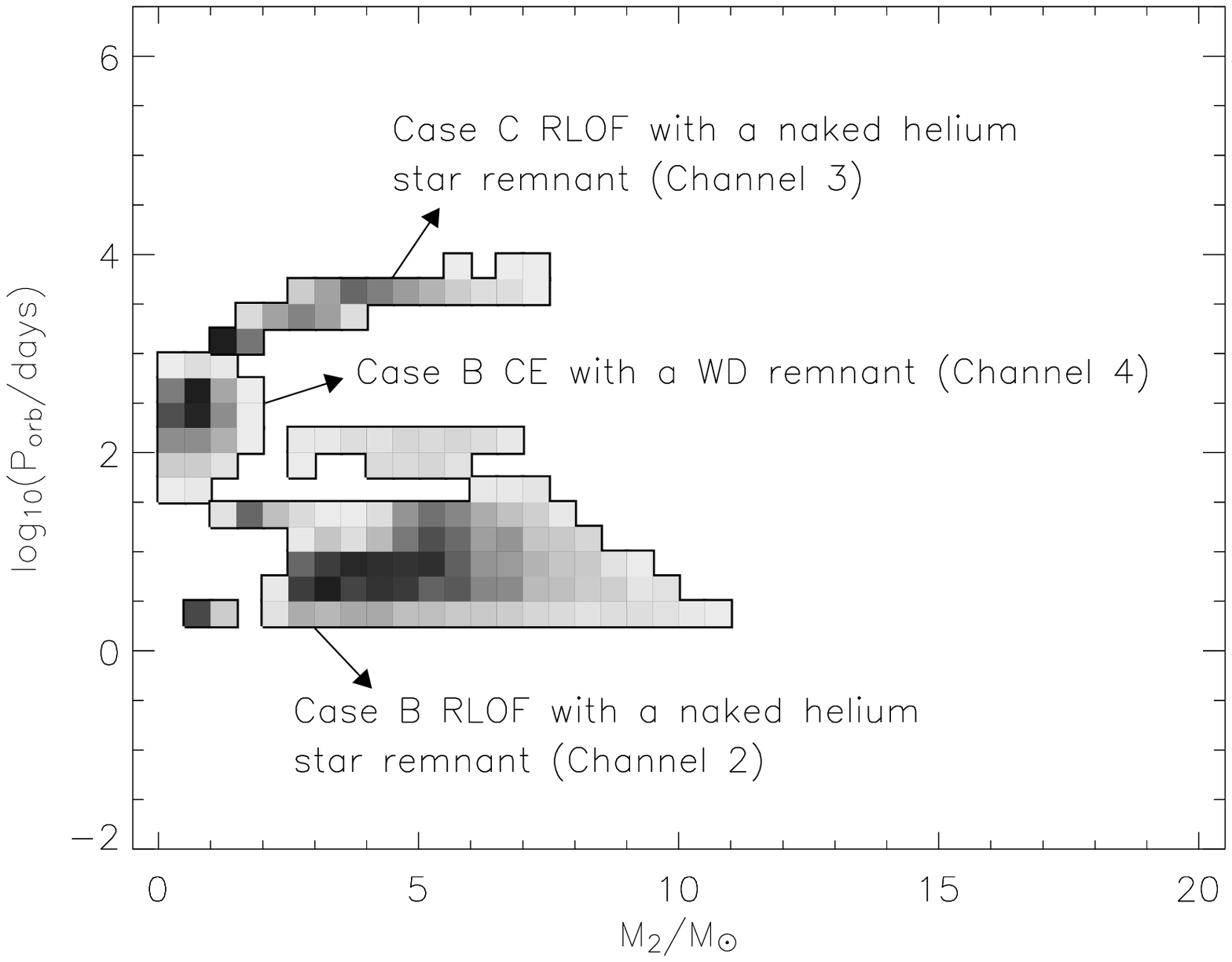}}
\resizebox{8.4cm}{!}{\includegraphics{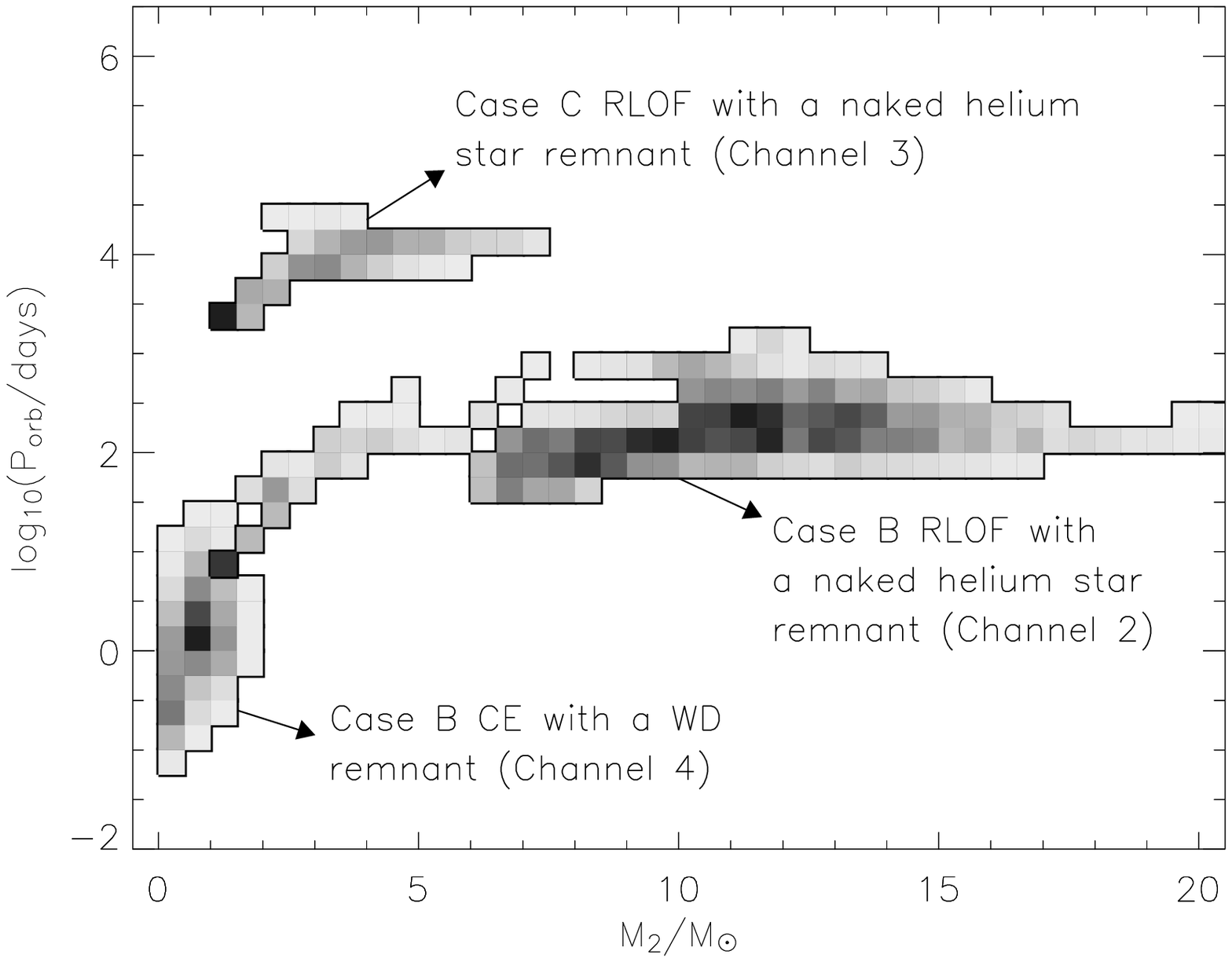}}
\resizebox{8.4cm}{!}{\includegraphics{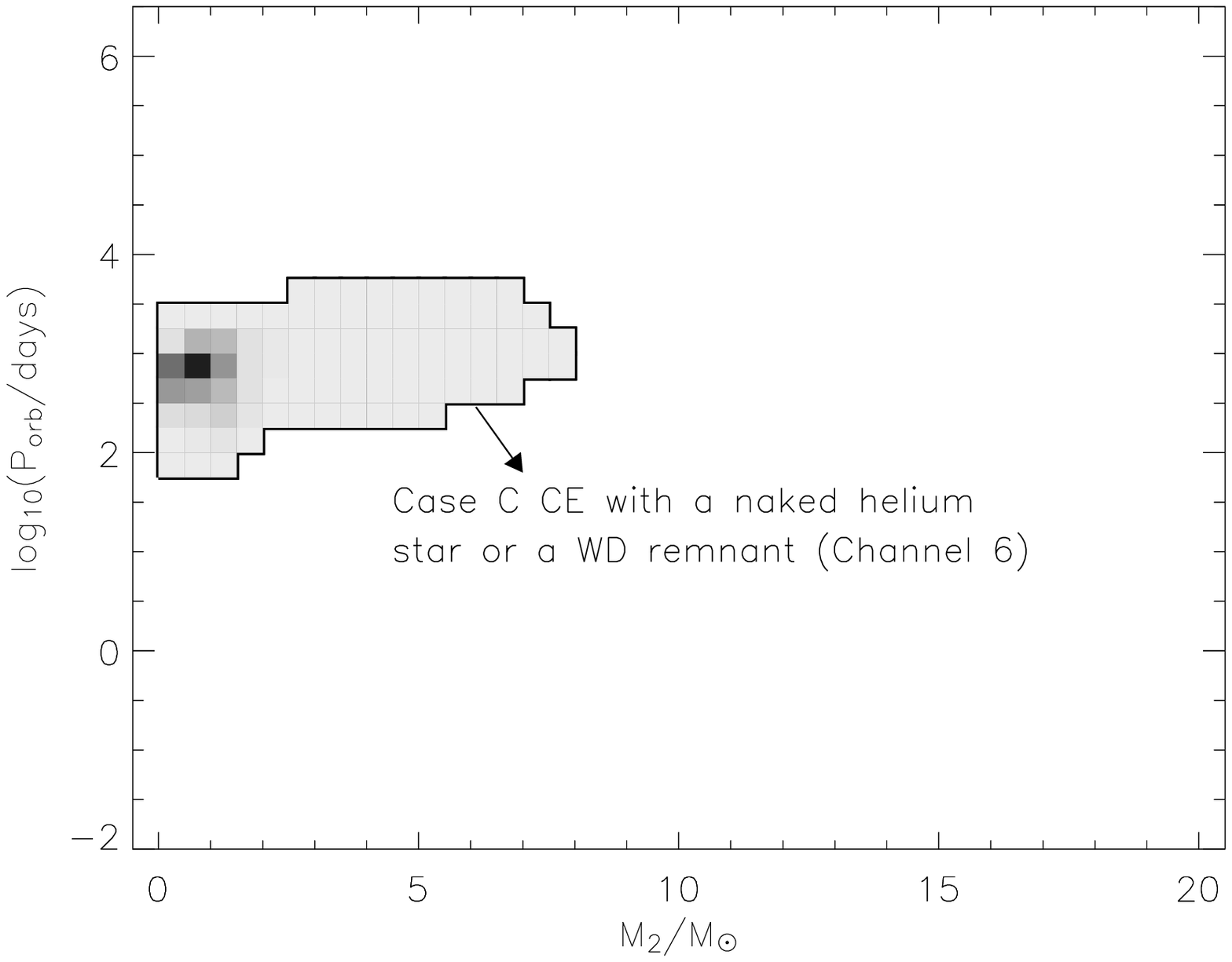}}
\resizebox{8.4cm}{!}{\includegraphics{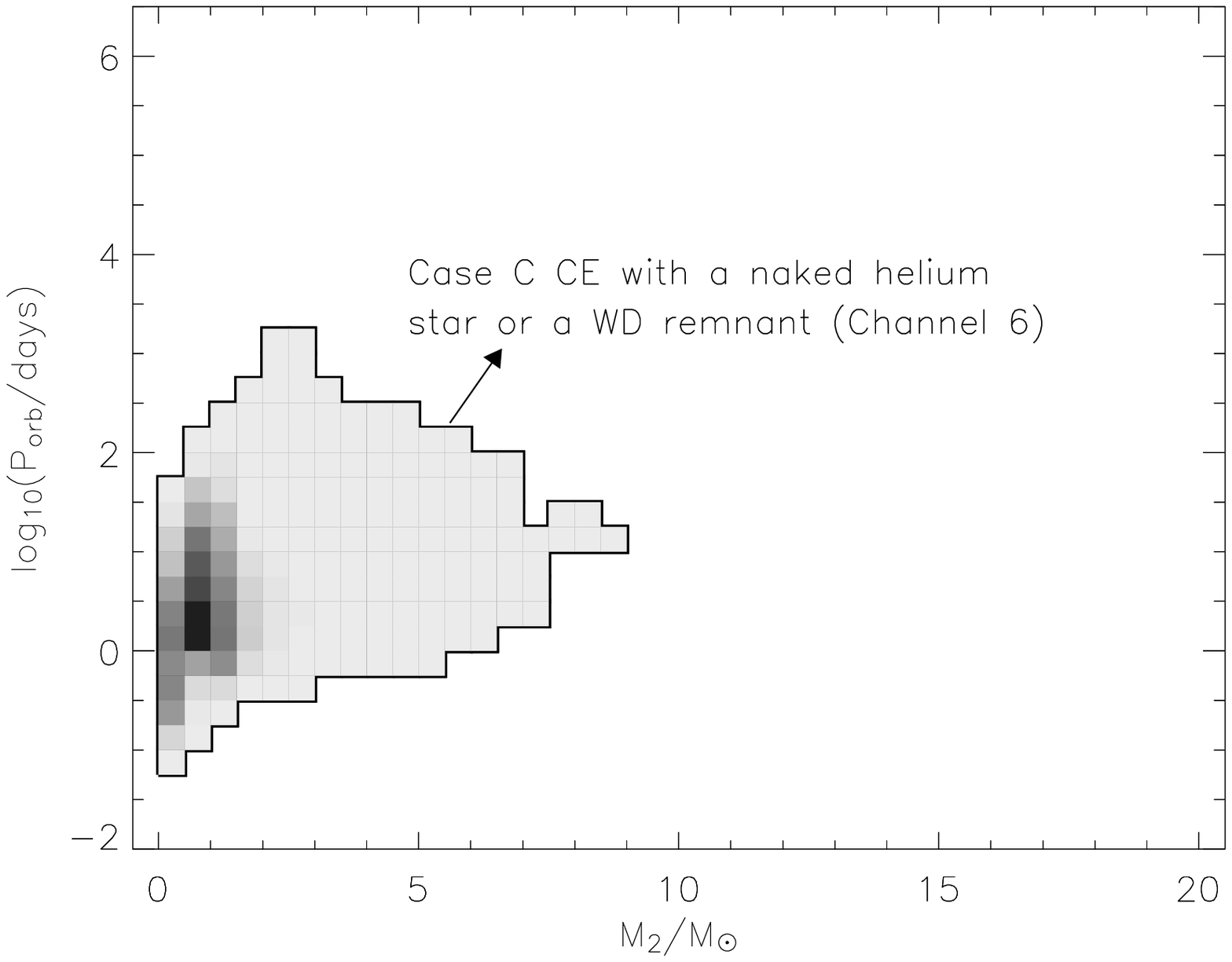}}
\caption{Distribution of present-day WDMS binaries in the $\left( M_2,
  P_{\rm orb} \right)$-plane at the beginning of the progenitors'
  evolution (left-hand panels) and at the beginning of the WDMS binary
  phase (right-hand panels). The distributions are normalised in the
  same way as in Fig.~\ref{2DM1}.}
\label{2DM2}
\end{center}
\end{figure*}

The two-dimensional probability distribution functions (PDFs)
describing the regions of the $\left( M_1, P_{\rm orb} \right)$-
and the $\left( M_2, P_{\rm orb} \right)$-planes occupied by binaries
evolving through formation channel~1 are displayed in the upper panels
of Figs.~\ref{2DM1} and~\ref{2DM2}. The left-hand panels of the figures
show the distributions of the WDMS binary progenitors at the beginning
of their evolution as {\em zero-age main-sequence binaries}, while the
right-hand panels show the distributions of the WDMS binaries {\em at 
the time of their formation}. For the construction of the plots, only
binaries contributing to the present-day Galactic population are taken
into account\footnote{Note that our purpose here is to examine the
formation space of WDMS binaries. The plots are therefore not fully
representative for the present-day population of short-period
systems ($P_{\rm orb} \la 1$ day) which may have undergone significant
orbital shrinkage due to magnetic braking and/or gravitational
radiation after their formation. However, even though the orbital
evolution is not shown in the figures, the finite life time of the
systems resulting from the orbital evolution and/or the nuclear
evolution of the secondary is taken into account in the
determination of the WDMS binaries currently
populating the Galaxy.}. The distributions are  
normalised so that the bin containing the largest contribution to the
population has a PDF value equal to one. In order to show the maximum
amount of detail this normalisation is performed separately for each
of the formation channels considered. The relative importance of the
different formation channels will be illustrated in Section~\ref{full}
and addressed in more detail in Section~\ref{numbers}. 

The initial orbital periods of the binaries evolving through formation
channel~1 typically range from 0.5 to 3 days. Binaries with shorter
orbital periods tend to converge before the primary is able to evolve
into a white dwarf due to the angular momentum losses caused by
magnetic braking and/or gravitational radiation (Pylyser \& Savonije
1988, 1989). Systems with longer orbital periods on the other hand
have somewhat too evolved donor stars to initiate a dynamically stable
Roche-lobe overflow phase. Instead, they undergo a common envelope
phase which, in view of the short initial periods, generally results
in the merger of the donor star's core with its main-sequence
companion.

The initial primary and secondary masses available to formation
channel~1 are limited to the intervals given by $1\,M_\odot \la M_1
\la 5\,M_\odot$ and $M_2 \la 3\,M_\odot$. The primary mass interval
arises from the dual requirement that the primary must be massive
enough to evolve away from the zero-age main sequence within the
imposed age limit of 10\,Gyr, but not so massive that its core mass at
the end of the mass-transfer phase is high enough to ignite helium in
its central layers. In our model, stars more massive than $\sim
2.5\,M_\odot$ are still able to evolve into helium white dwarfs due to
the reduction in mass caused by the thermal time scale mass-transfer
phase on the main sequence.  The upper limit of $3\,M_\odot$ on the
secondary mass is imposed by the secondary's main-sequence life time
which needs to be long enough to allow the primary to evolve into a
white dwarf before the secondary leaves the main sequence. Since the
life time of a star on the main sequence decreases with increasing
values of its mass, conservative mass transfer here yields a
competitive race between the formation of the white dwarf and the
accelerating evolution of the secondary. A similar behaviour can be
deduced from the evolutionary scenarios described by Iben \& Tutukov
(1985).

The formation channel eventually gives rise to WDMS binaries
consisting of a $0.1-0.4\,M_\odot$ He white dwarf and a main-sequence
star with a mass up to $6\,M_\odot$. Due to the stable mass-transfer
phase on the giant branch the orbital periods may be as long as 100
days.  The majority of the newly formed WDMS binaries have an orbital
period which is correlated with the mass of the white dwarf. The
correlation arises during the mass-transfer phase on the giant branch
where the radius of the giant, which is approximately equal to the
radius of its Roche-lobe, is determined by the mass of its core. The
same relation gives rise to the well-known correlation between the
white dwarf mass and the orbital period in wide binary millisecond
pulsars (see, e.g., Joss et al. 1987, Savonije 1987, Rappaport et
al. 1995, Ritter 1999, Tauris \& Savonije 1999). The small number of
WDMS binaries occupying the region below the $P_{\rm orb} - M_{\rm
WD}$ relation correspond to thermal time scale mass-transfer systems
for which the primary already lost most its envelope prior to the
stable mass-transfer phase on the giant branch.  The orbital periods
of the newly formed WDMS binaries furthermore increase with increasing
mass of the secondary. This relation arises from the narrow range of
initial orbital periods and secondary masses available to the
formation channel and from the increase of the orbital period with the
amount of mass transferred during the conservative Roche-lobe overflow
phase on the giant branch.

\subsubsection{Case B RLOF with a naked helium star remnant}
\label{sch2}

\begin{figure}
\resizebox{8.4cm}{!}{\includegraphics{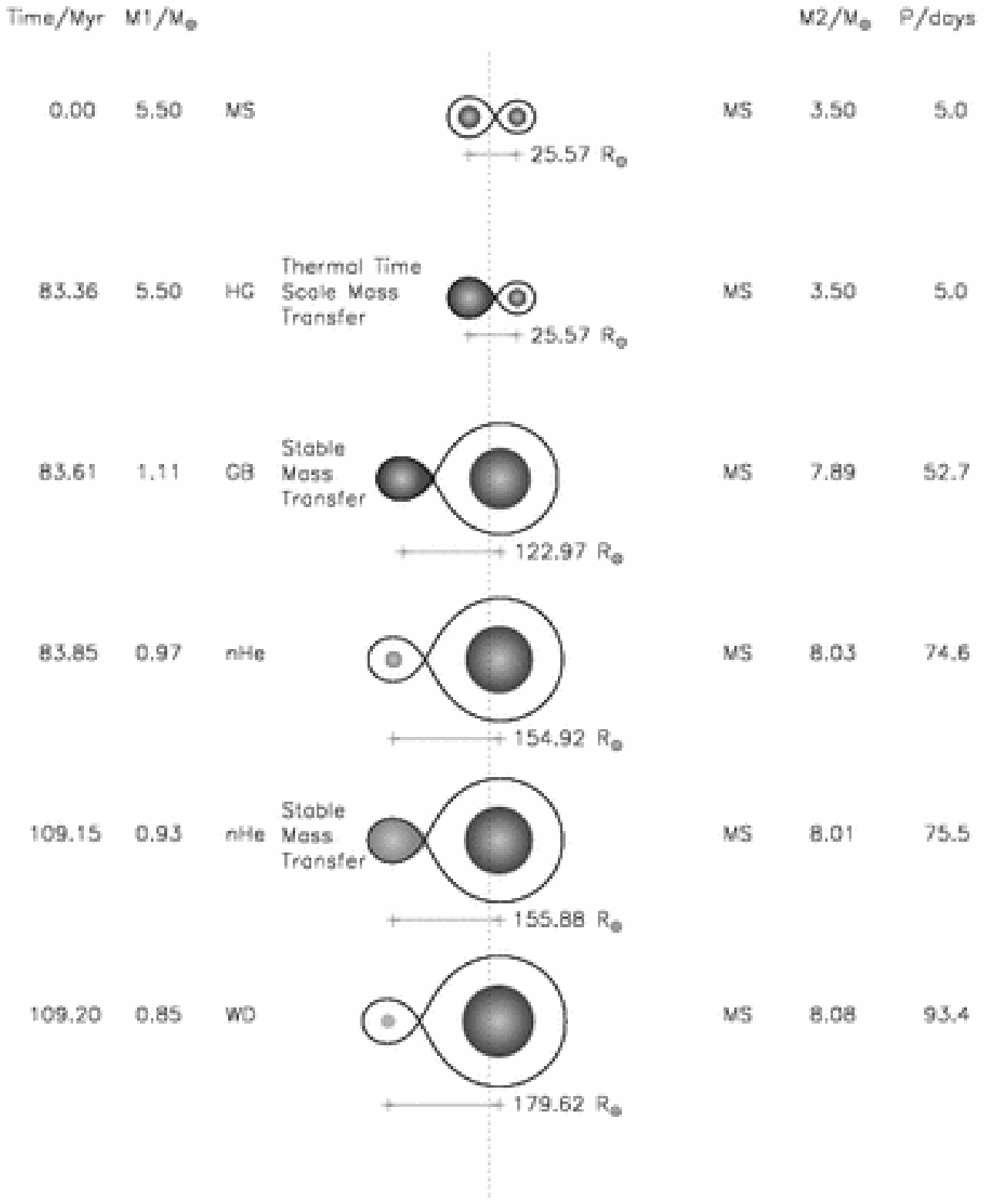}}
\caption{As Fig.~\ref{ch1}, but for formation channel~2. nHe
  stands for naked helium star.}
\label{ch2}
\end{figure}

Binaries evolving through the second formation channel start their
evolution with more massive component stars and somewhat longer
orbital periods than those in channel~1. They again undergo a highly
conservative case~B mass-transfer phase, possibly preceded by a short
case~A phase during the final stages of the primary's evolution on the
main sequence. Since the initial primary masses are higher and the
initial orbital periods are longer than in channel~1, the core of the
Roche-lobe filling giant is now massive enough to ignite helium in its
central layers so that a naked helium star is formed instead of a
white dwarf. During the subsequent evolution, the naked helium star at
some point loses its envelope either through the action of a stellar
wind or through a short second phase of stable Roche-lobe
overflow. The outcome in both cases is a relatively wide binary
consisting of a C/O or O/Ne/Mg white dwarf and an intermediate- to
high-mass main-sequence secondary. The main evolutionary phases of
this formation channel are summarised schematically in
Fig.~\ref{ch2}. We will refer to the channel as channel~2.

From the middle panels of Figs.~\ref{2DM1} and~\ref{2DM2}, it follows
that most of the WDMS binaries evolving through formation channel~2
originate from binaries with initial orbital periods between 2 and 130
days, initial primary masses between $2\,M_\odot$ and $12\,M_\odot$,
and initial secondary masses smaller than $11\,M_\odot$.  We note that
there is a small number of systems with primary masses up to
$20\,M_\odot$ for which highly conservative mass transfer results in
the formation of secondaries with masses up to $30\,M_\odot$. However,
since it is uncertain whether or not a white dwarf may be formed from
primaries with such high initial masses, we do not include these
systems in the following discussion.

The limits on the range of initial orbital periods result from the
evolutionary stage of the primary at the onset of the first Roche-lobe
overflow phase. For binaries with initial orbital periods shorter than
2 days, the core of the primary emerges from the mass-transfer phase
as a low-mass naked helium star with a helium-burning life time that is
too long for it to form a white dwarf before the secondary leaves the
main sequence. Binaries with initial orbital periods longer than 130
days on the other hand have significantly evolved donor stars with
deep convective envelopes. They are therefore subjected to a
common-envelope phase instead of to a thermal time scale mass-transfer
phase.

The lower limit of $2\,M_\odot$ on the mass of the primary stems from
the requirement that the star must be able to develop a sufficiently
massive core capable of igniting helium at the end of the stable
mass-transfer phase on the giant branch. The upper limit of
$12\,M_\odot$ corresponds to the highest mass for which a star
subjected to mass loss may evolve into a white dwarf rather than into
a neutron star (e.g. van den Heuvel 1981, Law \& Ritter 1983). The
lower and upper limits on the secondary mass result from the
requirement that mass transfer from the primary is dynamically stable
and from our convention that the secondary is initially less massive
than the primary. 

At the time of formation, the WDMS binaries forming through formation
channel~2 typically consist of a $0.65-1.44\,M_\odot$ C/O or O/Ne/Mg
white dwarf and a $1-20\,M_\odot$ main-sequence star. The orbital
periods range from 10 to 1000 days, with the bulk of the systems
occupying a rather narrow band of orbital periods around $P_{\rm orb}
\approx 200$ days.

\subsubsection{Case C RLOF with a white dwarf remnant}
\label{sch3}

\begin{figure}
\resizebox{8.4cm}{!}{\includegraphics{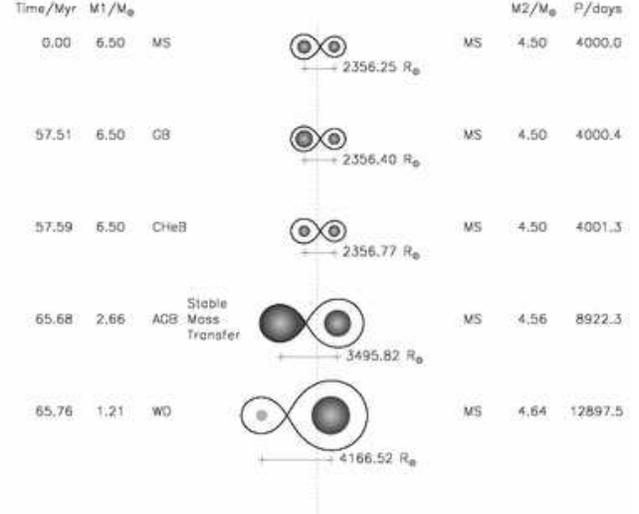}}
\caption{As Fig.~\ref{ch1}, but for formation channel~3. CHeB
  stands for core helium burning on the horizontal branch.}
\label{ch3}
\end{figure}

The third evolutionary channel applies to binaries with initial orbits
wide enough to allow the primary to evolve to the bottom of the
asymptotic giant branch (AGB) without overflowing its Roche lobe. As
it ascends the AGB, a strong stellar wind decreases the primary's mass
below 2/3 of the mass of its companion so that when it finally does
fill its Roche-lobe, the ensuing mass-transfer phase is dynamically
stable (e.g. Webbink et al. 1983). When mass transfer
ends, any remaining surface layers are 
quickly removed by the wind, leading to the exposure of the AGB star's
C/O or O/Ne/Mg core as a white dwarf. The orbital period of the newly
formed WDMS binary may be substantially longer than the initial binary
period due to the combined effect of the stellar wind and the stable
mass-transfer phase on the AGB. The main evolutionary phases followed
by a binary evolving through this formation channel are summarised
schematically in Fig.~\ref{ch3}. We will refer to the channel as
formation channel~3.

The middle panels of Figs.~\ref{2DM1} and~\ref{2DM2} show that the
binaries evolving through formation channel~3 have initial orbital
periods between 600 and 10\,000 days. In binaries with shorter
orbital periods the AGB wind has insufficient time to decrease the
mass ratio below 2/3 before the primary fills its Roche lobe. The
resulting mass-transfer phase is then dynamically unstable and leads
to the formation of a common envelope (see
Section~\ref{sch6}). Binaries with initial orbital periods longer than
10\,000 days are too wide to interact.

The initial primary and secondary masses range from $1\,M_\odot$ to
$8\,M_\odot$, similar to the mass range leading to the formation of
white dwarfs by single stars. This conformity arises because the
primary's Roche-lobe overflow phase does not take place until the very
late stages of the AGB (see also Iben \& Tutukov 1985). The fact that
the mass ratio needs to drop below 2/3 furthermore implies that the
initial secondary mass generally cannot be too much smaller than the
initial primary mass.

The formation channel eventually gives rise to WDMS binaries
consisting of a $0.5-1.44\,M_\odot$ C/O or O/Ne/Mg white dwarf and a
$1-8\,M_\odot$ main-sequence star revolving around each other with a
period of 2000 to 20000 days.

\subsection{Dynamically unstable mass transfer}
\label{unstable}

\subsubsection{Case B CE phase with a white dwarf remnant}

\begin{figure}
\resizebox{8.4cm}{!}{\includegraphics{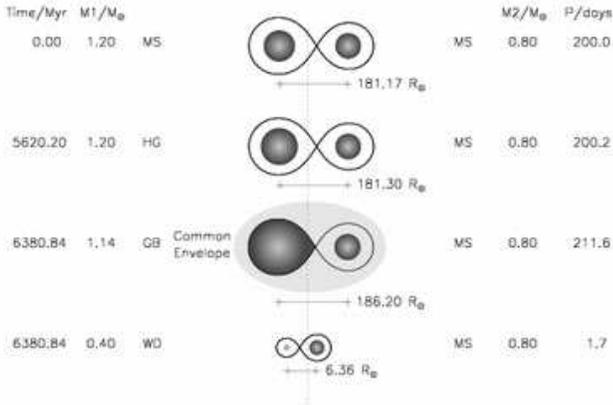}}
\caption{As Fig.~\ref{ch1}, but for formation channel~4.}
\label{ch4}
\end{figure}

The fourth evolutionary channel is characterised by a dynamically
unstable mass-transfer phase from a low-mass giant-branch star. As the
secondary plunges into the donor star's rapidly expanding envelope,
the orbit shrinks and orbital energy is transferred to the envelope
until it is expelled from the system. At the end of the phase, the
core of the Roche-lobe filling giant is exposed as a helium white
dwarf which orbits the main-sequence secondary with a drastically
reduced orbital separation. The main evolutionary phases comprising
this formation channel are summarised schematically in
Fig.~\ref{ch4}. We will refer to the channel as channel~4.

It follows from the middle panels of Figs.~\ref{2DM1} and~\ref{2DM2}
that the systems evolving through formation channel~4 have initial
orbital periods between 30 and 1000 days. Binaries with initial
orbital periods shorter than 30 days do not survive the common-envelope
phase evoked by the primary, while binaries with orbital periods
longer than 1000 days do not undergo Roche-lobe overflow until the
primary reaches the AGB. The limits on the initial primary and
secondary mass ranges, $1\,M_\odot \la M_1 \la 2\,M_\odot$ and $M_2
\la 2\,M_\odot$, arise for similar reasons as those in formation
channel~1 (see Section~\ref{sch1}). The upper limits on $M_1$ and $M_2$
are here somewhat smaller because the longer initial orbital periods
allow a primary of a given mass to reach a more evolved evolutionary
state and thus to develop a more massive helium core than the same
primary in a shorter-period binary. The mass and period ranges of
binaries surviving the common-envelope phase in this evolutionary
channel are in excellent agreement with the three-dimensional
hydrodynamical simulations performed by Sandquist et al. (2000). 

The formation channel results in WDMS binaries consisting of a
$0.3-0.5\,M_\odot$ He white dwarf and a main-sequence star with a mass
up to $2\,M_\odot$. The orbital period takes values between 0.05 and
30 days. The majority of the systems have $P_{\rm orb} \la 3$ days,
$M_{\rm WD} \approx 0.4\,M_\odot$  and $M_2 \la 1\,M_\odot$.

\subsubsection{Case B CE phase with a naked helium star remnant}
\label{sch5}

\begin{figure}
\resizebox{8.4cm}{!}{\includegraphics{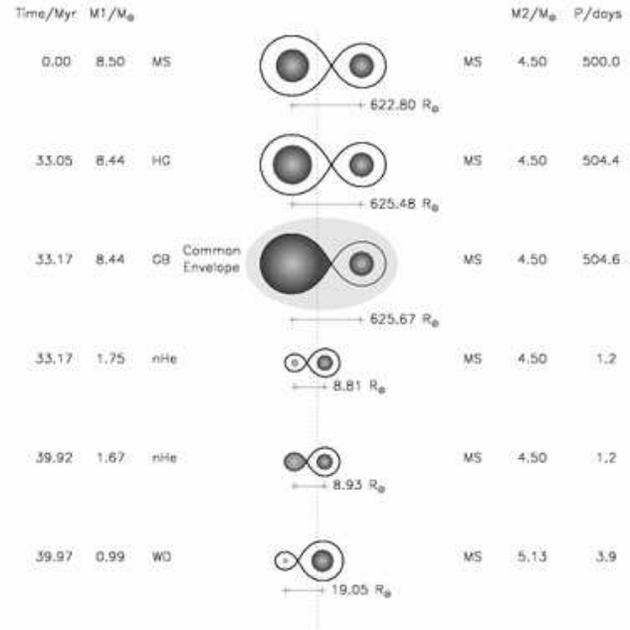}}
\caption{As Fig.~\ref{ch1}, but for formation channel~5.}
\label{ch5}
\end{figure}

In the fifth evolutionary channel, a common-envelope phase occurs when
an intermediate- to high-mass primary crosses the Hertzsprung-gap or
ascends the first giant branch. The core of the primary emerges from
the spiral-in phase as a low-mass naked helium star which orbits the
main-sequence secondary with an orbital period of the order of a few
days. The following few million years, the primary burns helium in its
core until it loses its surface layers in a stellar wind or until it
overflows its Roche lobe a second time when helium is exhausted
in the core. In either case, the primary evolves into a C/O or O/Ne/Mg
white dwarf orbiting an intermediate- to high-mass main-sequence
secondary. The main evolutionary phases characterising this formation
channel are summarised schematically in Fig.~\ref{ch5}.  We will refer
to this channel as formation channel~5.

The upper panels of Figs.~\ref{2DM1} and~\ref{2DM2} show that the
formation channel applies to binaries with initial orbital periods
between 10 and 1000 days, initial primary masses between $4\,M_\odot$
and $12\,M_\odot$, and initial secondary masses between $1\,M_\odot$
and $11\,M_\odot$. Stars in binaries with orbital periods shorter then
10 days are so close that Roche-lobe overflow from the primary usually
already occurs before it has developed a deep convective envelope, so
that mass transfer tends to be dynamically stable (see formation
channel~2, Section~\ref{sch2}). Binaries with periods longer than 1000
days on the other hand allow the primary to evolve beyond the first
giant branch without filling its Roche lobe. The gap in the initial
period range around $\sim 100$ days separates systems initiating mass
transfer in the Hertzsprung gap from those initiating mass transfer on
the giant branch. Mass transfer from Hertzsprung-gap donor stars often
starts of as a thermal time scale mass-transfer phase which evolves
into a common-envelope phase as the donor star approaches the giant
branch. The survival of these systems depends strongly on the ability
of the thermal time scale mass-transfer phase to decrease the
primary's mass sufficiently before the onset of the common-envelope
phase. The gap between systems with Hertzsprung-gap and giant-branch
donor stars is related to the behaviour of the adiabatic radius-mass
exponents tabulated by Hjellming (1989). The large values found by
Hjellming (1989) near the transition phase where the star starts to
develop a deep convective envelope yield a small window in the
parameter space where the mass-transfer phase is dynamically stable,
so that a different evolutionary scenario ensues (e.g. formation
channel~2). However, in view of the still existing uncertainties in
the detailed modelling of this transition phase, the associated values
of the adiabatic radius-mass exponents are also quite uncertain. 
The occurrence of the gap may therefore be an artifact of the
stability criterion separating systems undergoing dynamically stable
Roche-lobe overflow from those undergoing dynamically unstable
Roche-lobe overflow. The origin of the limits on the primary mass
range is similar to that of the limits found for formation channel~2
(Section~\ref{sch2}). The lower limit of $1\,M_\odot$ on the secondary
mass corresponds to the smallest companion mass for which the binary
is able to avoid a merger.

At the time of formation, the WDMS binaries forming through formation
channel~5 consist of a $0.65-1.44\,M_\odot$ C/O or O/Ne/Mg white dwarf
and a $2-13\,M_\odot$ main-sequence star orbiting each other with a
period of 0.5 to 20 days. The high secondary masses result
from mass accretion during the second Roche-lobe overflow phase of the
primary.

\subsubsection{Case C CE phase with a naked helium star or a white
  dwarf remnant}
\label{sch6}

\begin{figure}
\resizebox{8.4cm}{!}{\includegraphics{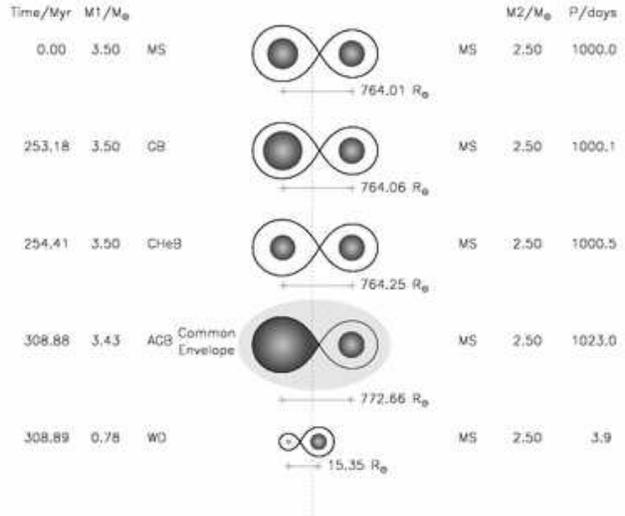}}
\caption{As Fig.~\ref{ch1}, but for formation channel~6.}
\label{ch6}
\end{figure}

Similar to formation channel~3, the initial orbital separations of the
binaries following the sixth evolutionary channel are wide enough to
avoid any type of Roche-lobe overflow until the primary reaches the
AGB. However, the primary here fills its critical Roche lobe before
the wind has a chance to reduce the primary's mass below 2/3 that of
its companion so that the resulting mass-transfer phase is now
dynamically unstable. The binary emerges from the ensuing
common-envelope phase as a WDMS binary consisting of a C/O or O/Ne/Mg
white dwarf, the former core of the AGB primary, and a low- to
intermediate-mass main-sequence star. If the common-envelope phase
takes place early on the AGB, the C/O or O/Ne/Mg core may retain a
thin helium envelope which is subsequently stripped away by a
Wolf-Rayet type stellar wind.  The main evolutionary phases occurring
in the formation channel are summarised schematically in
Fig.~\ref{ch6}. We will refer to this channel as formation channel~6.

From the bottom panels of Figs.~\ref{2DM1} and~\ref{2DM2} it follows
that the systems evolving through formation channel~6 originate from
binaries with initial orbital periods between 60 and 6500 days,
initial primary masses between $1\,M_\odot$ and $9\,M_\odot$, and
initial secondary masses smaller than or equal to $8\,M_\odot$. The
lower limit on the orbital period range stems from the requirement
that the binaries survive the common-envelope phase, while the upper
limit corresponds to the longest orbital period for which the primary
fills its Roche lobe on the AGB. The limits on the primary and
secondary masses arise for similar reasons as those in channel~3 (see
Section~\ref{sch3}).

The WDMS binaries forming through formation channel~6 typically
consist of a $0.5-1.44\,M_\odot$ C/O or O/Ne/Mg white dwarf and a
main-sequence star with a mass up to $8\,M_\odot$. The orbital period
ranges from 0.05 to 2000 days, where the long-period tail corresponds
to systems in which the envelope mass of the primary is negligible in
comparison to its core mass. A similar long-period tail was found by
de Kool \& Ritter (1993, see their Fig. 2a). Most WDMS binaries
forming through this channel have $M_{\rm WD} \approx
0.5-0.6\,M_\odot$, $M_2 \la 2\,M_\odot$, and $P_{\rm orb} \la 20$
days.

\subsection{Non-interacting systems}

The last formation channel is the most straightforward one as it
represents the non-interacting WDMS binaries. The initial orbital
periods of these systems are wide enough for the two stars to evolve
in much the same way as they would if they were single.  Their
contribution to the population is therefore independent of the
assumptions adopted for the treatment of mass transfer in
semi-detached binaries, so that they may provide a convenient means to
renormalise our results for comparison with observations and with
other authors.

The two-dimensional distribution functions describing the population
of WDMS binaries forming without interacting are displayed in the
upper panels of Figs.~\ref{2DM1} and~\ref{2DM2}. The initial orbital
periods are typically longer than 400 days, while the initial primary and
secondary masses range from $1\,M_\odot$ to $9\,M_\odot$ and from
$1\,M_\odot$ to $8\,M_\odot$, respectively. At the time of formation,
most binaries have $M_{\rm WD} \approx 0.5-0.6\,M_\odot$ and $M_2 \la
2\,M_\odot$. The final orbital periods may be substantially longer
than the initial ones due to the action the stellar wind responsible
for exposing the primary's core as a C/O or O/Ne/Mg white dwarf.

\subsection{The entire population}
\label{full}

\begin{figure*}
\begin{center}
\resizebox{12.6cm}{!}{\includegraphics{bwbar0to1.ps}} \\
\resizebox{8.4cm}{!}{\includegraphics{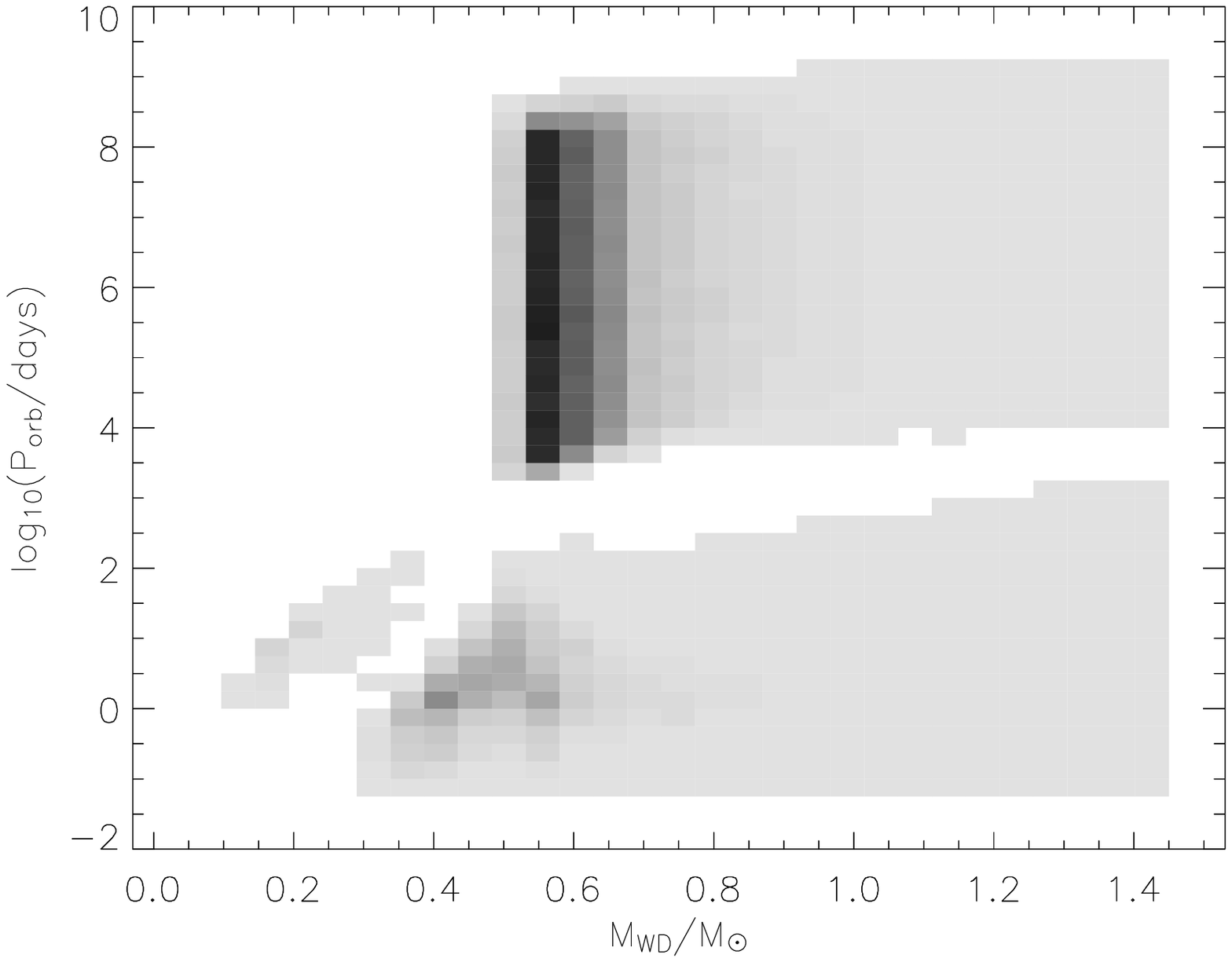}}
\resizebox{8.4cm}{!}{\includegraphics{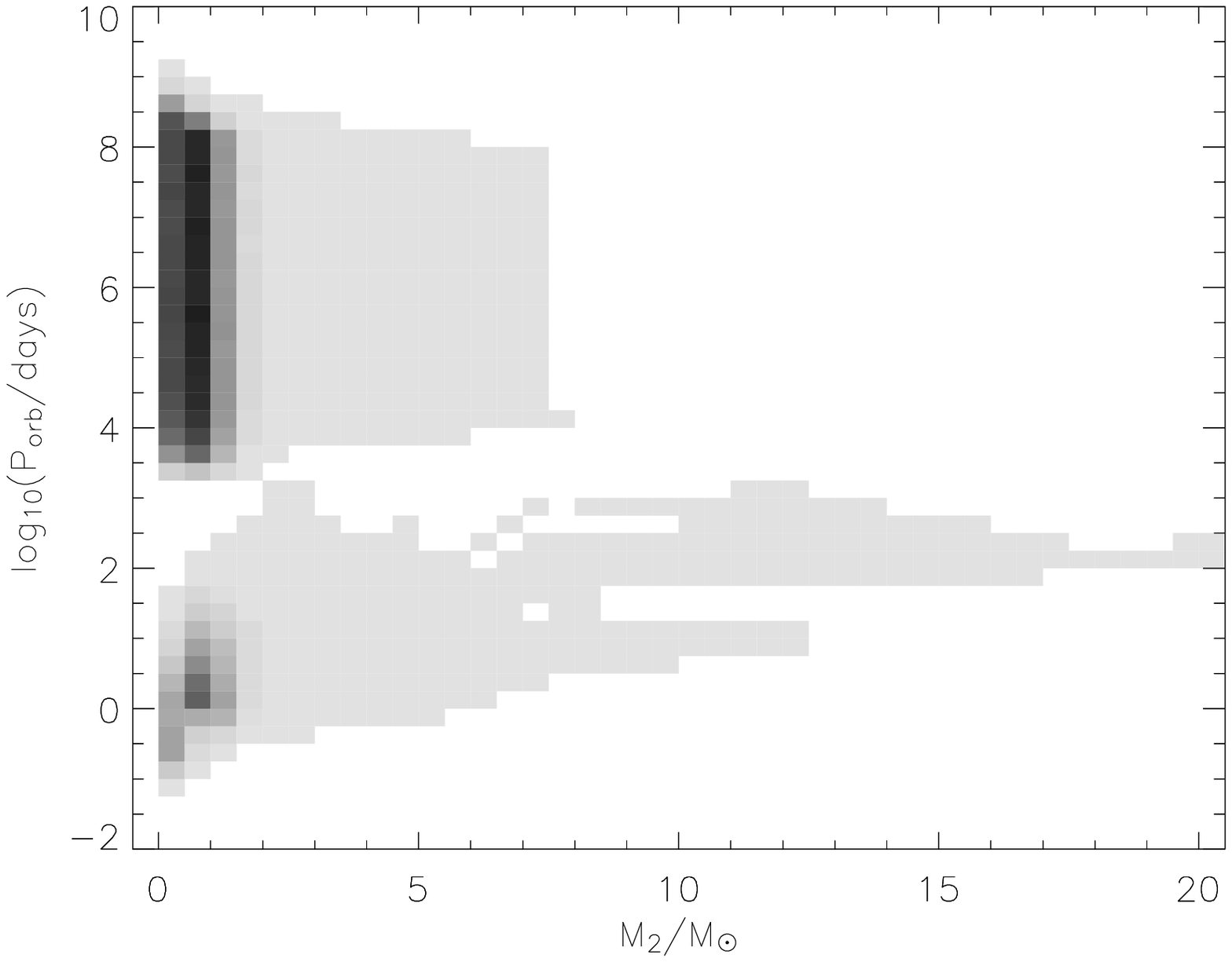}} \\
\caption{Distribution of present-day WDMS binaries in the 
  $\left( M_{\rm WD}, P_{\rm orb} \right)$- and $\left( M_2,
  P_{\rm orb} \right)$-planes at the start of the WDMS binary
  phase.}
\label{2Dfull}
\end{center}
\end{figure*}

\begin{figure*}
\begin{center}
\resizebox{5.9cm}{!}{\includegraphics{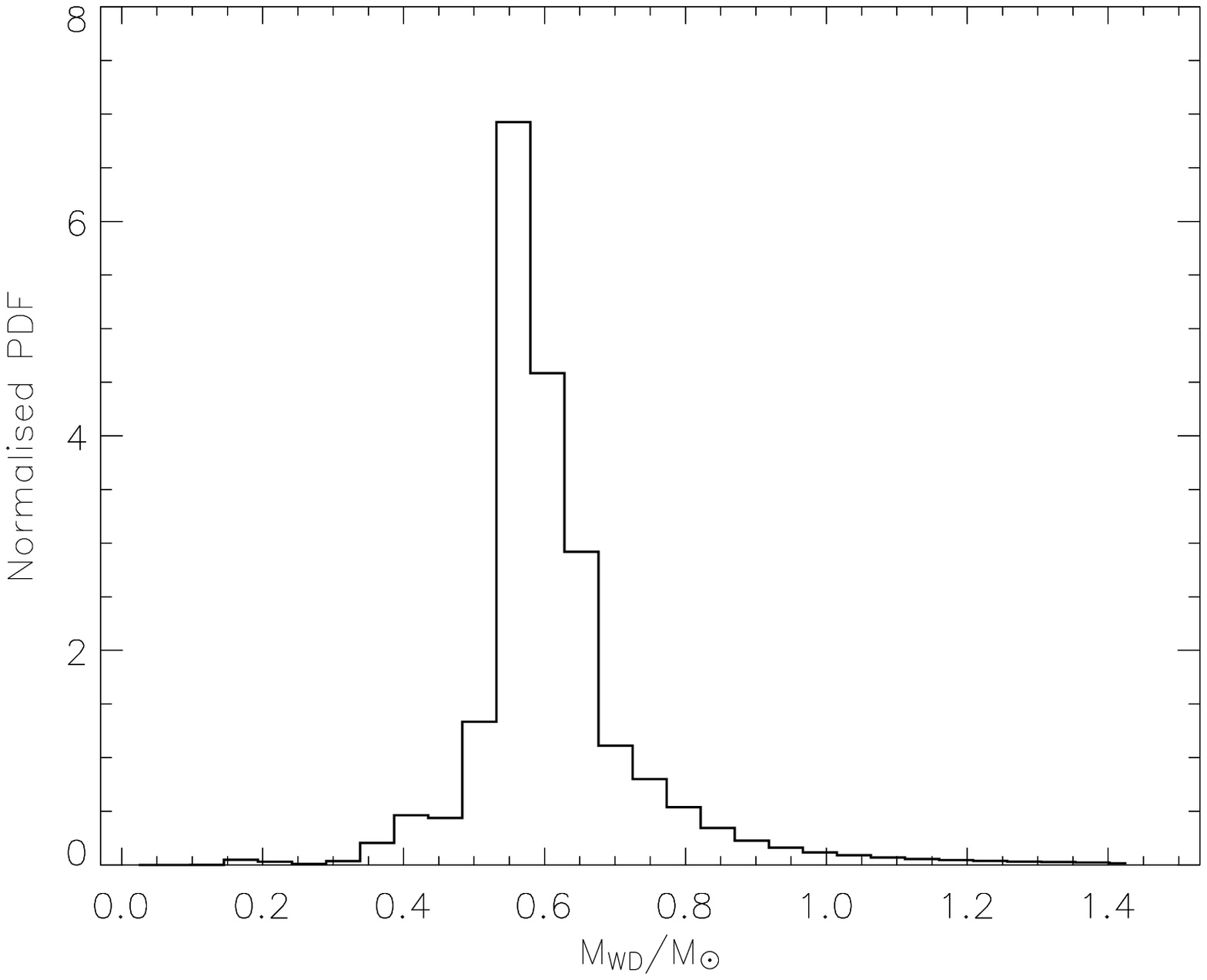}}
\resizebox{5.9cm}{!}{\includegraphics{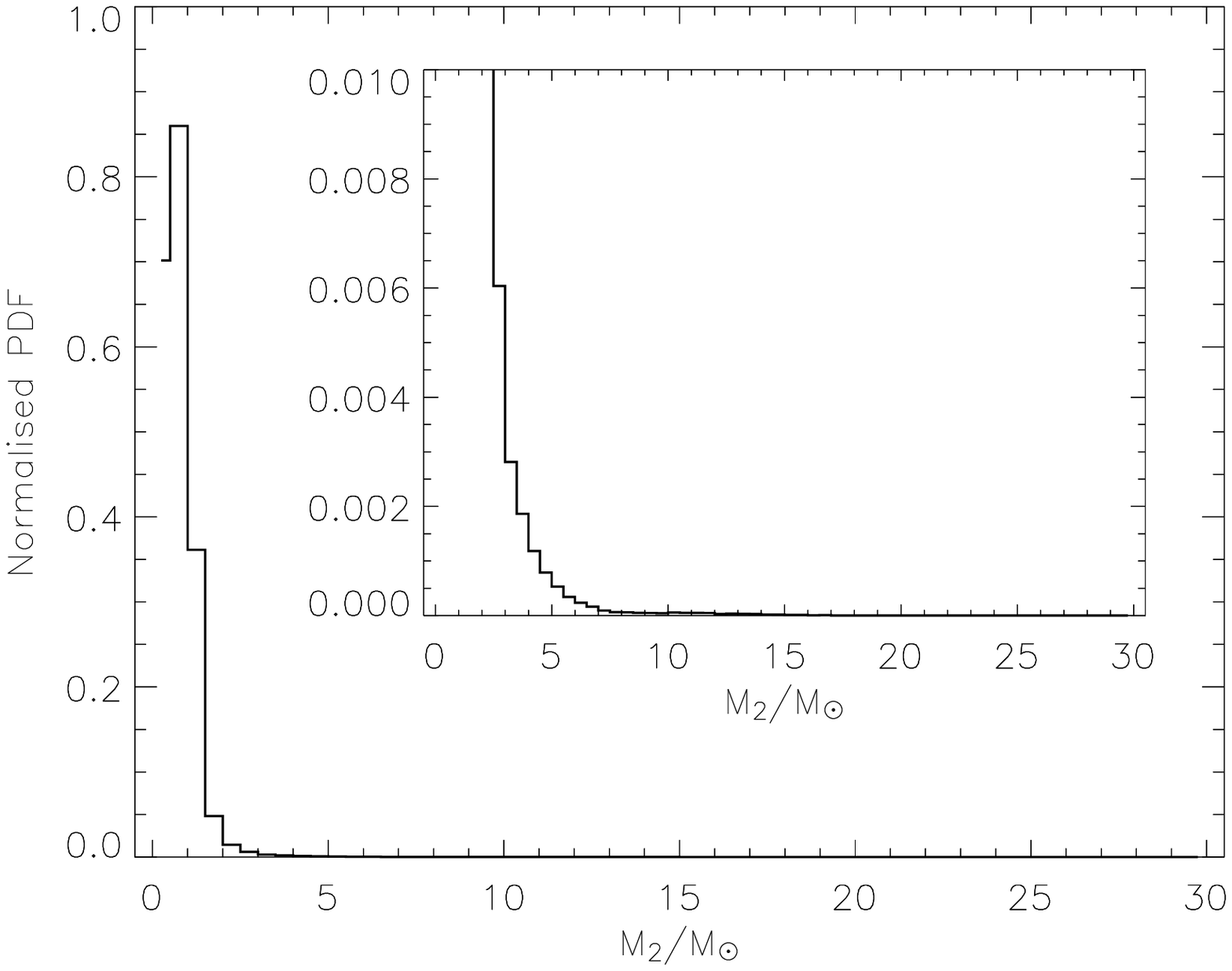}} 
\resizebox{5.9cm}{!}{\includegraphics{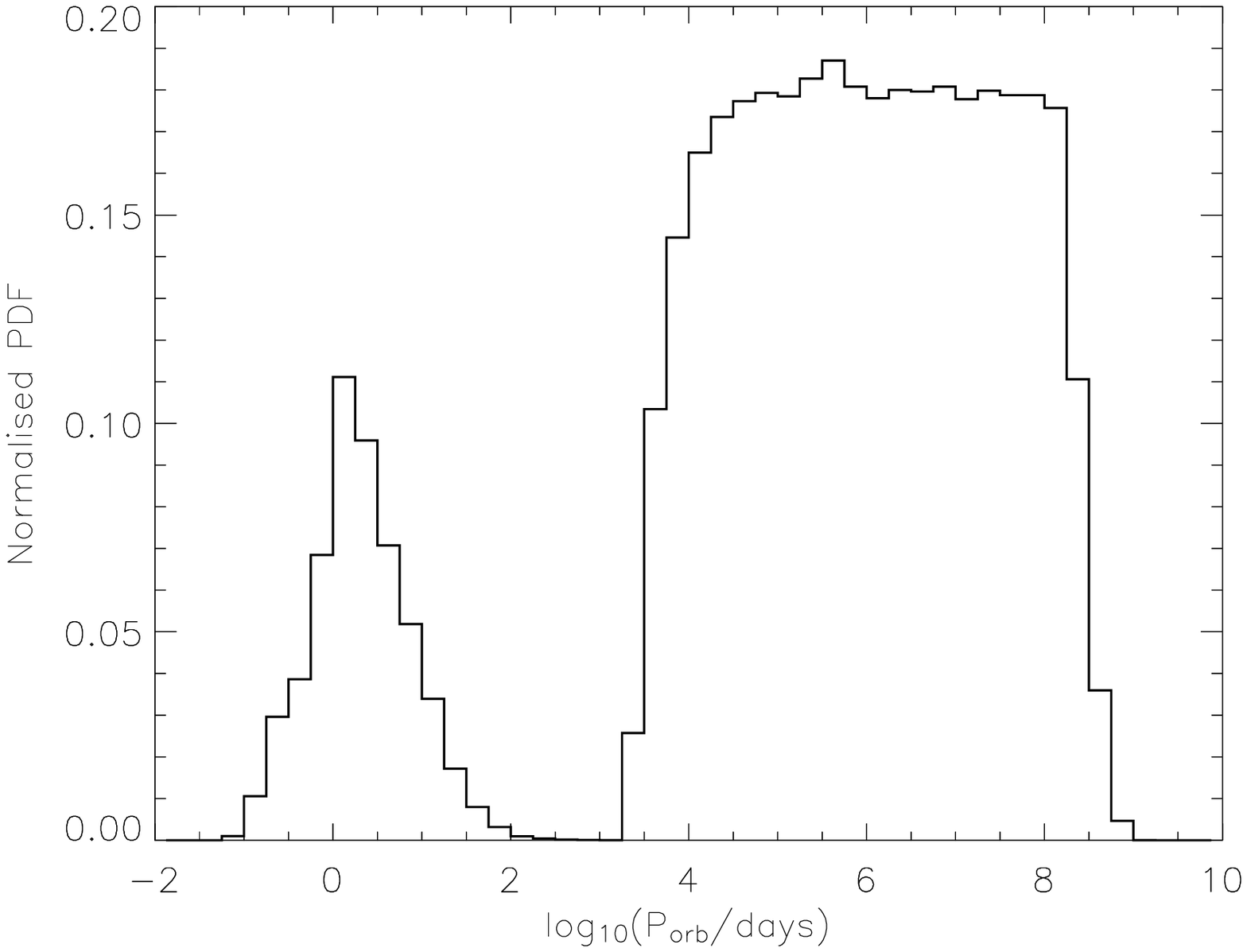}} 
\caption{White dwarf mass, secondary mass, and orbital period 
  distributions of present-day WDMS binaries at the start of the 
  WDMS binary phase. The distributions are normalised so that the
  integral over all systems found is equal to one.}
\label{1Dfull}
\end{center}
\end{figure*}

In order to get an idea of the relative importance of the different
formation channels, the formation space of all WDMS binaries forming
through channels 1--7 is shown in Fig.~\ref{2Dfull} without
renormalising the contributions of the different channels as in
Figs.~\ref{2DM1} and~\ref{2DM2}. The population is then clearly
dominated by wide non-interacting systems with low-mass main-sequence 
stars evolving through formation channel~7. The second largest
group consists of systems with low-mass main sequence stars and
periods in the range from 1 to 10 days originating from 
channels~4 and~6. The dominance of WDMS binaries with wide
non-interacting progenitors furthermore implies that the majority
of the systems contain a C/O white dwarf with a mass around $\sim
0.6\,M_\odot$. Systems with white dwarf masses higher than $\sim
0.8\,M_\odot$ or secondary masses higher than $2\,M_\odot$ are 
relatively rare. This is further illustrated by the
one-dimensional distribution functions for $M_{\rm WD}$, $M_2$, and
$P_{\rm orb}$ displayed in Fig.~\ref{1Dfull}. The bimodal nature of
the orbital period distribution is in good agreement with Fig. 5
of de Kool \& Ritter (1993). 

We will quantify the relative contributions of the different
formation channels in more detail in Section~\ref{numbers}.

\section{The effect of $\gamma_{\rm RLOF}$ and $\alpha_{\rm CE}$}
\label{sparam} 

The properties of the population of WDMS binaries forming through the
evolutionary channels described in Section~\ref{form} may depend
sensitively on the assumptions underlying the binary evolution
calculations. The two main uncertainties applying to the discussed
formation channels are the amount of mass accreted by a normal-type
star during episodes of stable Roche-lobe overflow and the treatment
of the common-envelope phase resulting from dynamically unstable
Roche-lobe overflow. In order to assess how these uncertainties affect
the WDMS binary population, we repeated the calculations presented in
Section~\ref{form} for different mass-accretion parameters
$\gamma_{\rm RLOF}$ and common-envelope ejection efficiencies
$\alpha_{\rm CE}$. The values of $\gamma_{\rm RLOF}$ and $\alpha_{\rm
CE}$ adopted in the different population synthesis models are listed in
Table~\ref{models}. Our standard model (model~A) corresponds to the
assumptions described and used in Sections~\ref{basic} and~\ref{form}.

\begin{table}
\caption{Population synthesis model parameters. }
\label{models}
\begin{tabular}{lcc}
\hline
\hline
 model & $\gamma_{\rm RLOF}$ & $\alpha_{\rm CE}$ \\
\hline
A    & Eq.~(\ref{gam}) & 1.0 \\
G0   & 0.0 & 1.0 \\
G025 & 0.25 & 1.0 \\
G05  & 0.5 & 1.0 \\
G075 & 0.75 & 1.0 \\
G1   & 1.0 & 1.0 \\
CE1  & Eq.~(\ref{gam}) & 0.2 \\
CE3  & Eq.~(\ref{gam}) & 0.6 \\
CE6  & Eq.~(\ref{gam}) & 3.0 \\
CE8  & Eq.~(\ref{gam}) & 5.0 \\
\hline
\end{tabular}
\end{table}

The effects of $\gamma_{\rm RLOF}$ and $\alpha_{\rm CE}$ on the
parameter space occupied by WDMS progenitors at the start of their
evolution and by WDMS binaries at the time of their formation are
summarised in Table~\ref{param}. As in Figs.~\ref{2DM1}
and~\ref{2DM2}, only binaries contributing to the present-day Galactic
population are taken into account.  In order to focus on the bulk of
the systems forming through each formation channel, the mass and
period ranges quoted are limited to bins containing at least $\sim$1\%
of the channel's contribution to the WDMS binary population. In the
case of model~A, the extent of the parameter space listed in the table
is therefore somewhat more restricted than in Section~\ref{form}. We
also note that by simply quoting the minimum and maximum values of
$M_1$, $M_2$, and $P_{\rm orb}$ we neglect the deviations of the
high-density regions in the parameter space from a simple
'rectangular' shape.

\begin{table*}
\caption{The effect of the population synthesis model parameters on
  the masses and orbital periods of WDMS binary progenitors at the
  start of their evolution and of WDMS binaries at the time of their
  formation. For each formation channel, the mass and period ranges
  refer to bins containing at least $\sim$1\% of the channel's
  contribution to the WDMS binary population. The quoted ranges should
  therefore be considered as approximate ranges for the bulk of the
  systems and not as strict limits. For the construction of the table,
  only binaries contributing to the present-day Galactic population
  are taken into account. Details of the parameters adopted in the
  different population synthesis models are listed in
  Table~\ref{models}.}
\label{param}
\begin{tabular}{cllcccccc}
\hline
\hline
 & & & \multicolumn{3}{c}{Initial Parameters} & \multicolumn{3}{c}{Formation Parameters} \\
\hline
Channel & \multicolumn{2}{l}{Model} & $M_1/M_\odot$ & $M_2/M_\odot$ & $P_{\rm orb}/{\rm days}$ & $M_{\rm WD}/M_\odot$ & $M_2/M_\odot$ & $P_{\rm orb}/{\rm days}$ \\
\hline
 1 & A   & $\!\!\!$(standard)                & $1-3$ & $<2$ & $0.5-3$  & $0.15-0.35$ & $<5$ & $1-100$  \\
   & G0  & $\!\!\!$($\gamma_{\rm RLOF}=0$)   & $1-3$ & $<2$ & $0.5-3$  & $0.15-0.35$ & $1-5$ & $3-100$  \\
   & G05 & $\!\!\!$($\gamma_{\rm RLOF}=0.5$) & $1-3$ & $<2$ & $0.5-3$  & $0.15-0.35$ & $<4$ & $2-100$  \\
   & G1  & $\!\!\!$($\gamma_{\rm RLOF}=1$)   & $1-3$ & $<2$ & $0.5-3$  & $0.15-0.35$ & $<3$ & $1-100$  \\
\hline
 2 & A   & $\!\!\!$(standard)               & $2-11$ & $1-9$ & $2-100$  & $0.3-1.44$ & $1-17$ & $10-1000$  \\
   & G0  & $\!\!\!$($\gamma_{\rm RLOF}=0$)  & $4-11$ & $2-9$ & $2-100$  & $0.7-1.44$ & $6-17$ & $30-1000$  \\
   & G05 & $\!\!\!$($\gamma_{\rm RLOF}=0.5$) & $2-10$ & $1-8$ & $2-100$ & $0.3-1.44$ & $2-11$ & $20-1000$  \\
   & G1  & $\!\!\!$($\gamma_{\rm RLOF}=1$)  & $2-9$  & $1-6$ & $2-100$  & $0.3-1.25$ & $1-6$ & $10-1000$  \\
\hline
 3 & A   & $\!\!\!$(standard)               & $1-8$ & $1-7$ & $600-6000$  & $0.55-1.44$ & $1-7$ & $2000-20000$  \\
   & G0  & $\!\!\!$($\gamma_{\rm RLOF}=0$)  & $1-8$ & $1-7$ & $600-6000$  & $0.55-1.44$ & $1-7$ & $2000-20000$  \\
   & G05 & $\!\!\!$($\gamma_{\rm RLOF}=0.5$) & $1-8$ & $<6$ & $600-6000$  & $0.5-1.4$  & $<6$  & $1000-20000$  \\
   & G1  & $\!\!\!$($\gamma_{\rm RLOF}=1$)  & $1-6$ & $<4$  & $200-6000$  & $0.5-0.95$  & $<4$  & $600-20000$  \\
\hline
 4 & CE1 & $\!\!\!$($\alpha_{\rm CE}=0.2$)  & $1-2$ & $<2$ & $200-600$ & $0.4-0.5$  & $<2$ & $0.1-3$  \\
   & A   & $\!\!\!$(standard)               & $1-2$ & $<2$ & $30-600$  & $0.3-0.5$  & $<2$ & $0.1-20$  \\
   & CE8 & $\!\!\!$($\alpha_{\rm CE}=5.0$)  & $1-2$ & $<2$ & $10-600$  & $0.25-0.5$ & $<2$ & $0.1-60$  \\
   & G0  & $\!\!\!$($\gamma_{\rm RLOF}=0$)  & $1-2$ & $<2$ & $30-600$  & $0.3-0.5$  & $<2$ & $0.1-20$  \\
   & G05 & $\!\!\!$($\gamma_{\rm RLOF}=0.5$) & $1-2$ & $<2$ & $30-600$ & $0.3-0.5$  & $<2$ & $0.1-20$  \\
   & G1  & $\!\!\!$($\gamma_{\rm RLOF}=1$)  & $1-2$ & $<2$ & $30-600$  & $0.3-0.5$  & $<2$ & $0.1-20$  \\
\hline
 5 & CE1 & $\!\!\!$($\alpha_{\rm CE}=0.2$)  & $8-11$ & $4-7$ & $30-100$   & $0.95-1.4$ & $7-11$ & $2-6$  \\
   & A   & $\!\!\!$(standard)               & $5-11$ & $2-9$ & $20-1000$  & $0.75-1.44$ & $3-11$ & $0.5-20$  \\
   & CE8 & $\!\!\!$($\alpha_{\rm CE}=5.0$)  & $3-11$ & $<7$  & $10-1000$   & $0.4-1.44$ & $<9$   & $0.3-30$  \\
   & G0  & $\!\!\!$($\gamma_{\rm RLOF}=0$)  & $6-11$ & $2-9$ & $20-1000$  & $0.8-1.44$ & $3-11$ & $1-20$  \\
   & G05 & $\!\!\!$($\gamma_{\rm RLOF}=0.5$) & $4-11$ & $2-9$ & $20-1000$ & $0.7-1.44$ & $2-10$ & $0.5-20$  \\
   & G1  & $\!\!\!$($\gamma_{\rm RLOF}=1$)  & $6-11$ & $2-9$ & $200-1000$ & $0.8-1.44$ & $2-9$  & $1-20$  \\
\hline
 6 & CE1 & $\!\!\!$($\alpha_{\rm CE}=0.2$)  & $1-2$ & $<3$ & $300-2000$  & $0.5-0.8$ & $<3$ & $0.1-10$  \\
   & A   & $\!\!\!$(standard)               & $1-5$ & $<3$ & $100-2000$  & $0.5-0.9$ & $<3$ & $0.1-60$  \\
   & CE8 & $\!\!\!$($\alpha_{\rm CE}=5.0$)  & $1-5$ & $<3$ & $30-2000$   & $0.5-0.9$ & $<3$ & $0.2-300$  \\
   & G0  & $\!\!\!$($\gamma_{\rm RLOF}=0$)  & $1-5$ & $<3$ & $100-2000$  & $0.5-0.9$ & $<3$ & $0.1-60$  \\
   & G05 & $\!\!\!$($\gamma_{\rm RLOF}=0.5$) & $1-5$ & $<3$ & $100-2000$ & $0.5-0.9$ & $<3$ & $0.1-60$  \\
   & G1  & $\!\!\!$($\gamma_{\rm RLOF}=1$)  & $1-5$ & $<3$ & $100-2000$  & $0.5-0.9$ & $<3$ & $0.1-30$  \\
\hline
 7 & A   & $\!\!\!$(standard)              & $1-5$ & $<2$ & $10^3-10^8$  & $0.5-0.9$ & $<2$ & $3 \times 10^3-5 \times 10^8$  \\
\hline
\end{tabular}
\end{table*}

Since the mass-accretion parameter $\gamma_{\rm RLOF}$ determines the
amount of mass and thus the amount of orbital angular momentum lost
from the system during episodes of dynamically stable Roche-lobe
overflow, the parameter directly affects the evolution of WDMS
binaries forming through formation channels 1--3. The parameter also
indirectly affects the formation of WDMS binaries through channels
4--6 because it enters the stability criterion separating dynamically
stable from dynamically unstable Roche-lobe overflow systems (for
details see, e.g., the appendix in Kolb et al. 2001). The dependency
of the criterion on $\gamma_{\rm RLOF}$ is such than an increase in
$\gamma_{\rm RLOF}$ generally increases the critical mass ratio
$Q_{\rm c}=M_{\rm donor}/M_{\rm accretor}$ separating dynamically
stable ($Q < Q_{\rm c}$) from dynamically unstable ($Q > Q_{\rm c}$)
Roche-lobe overflowing systems\footnote{Note that the definition of 
the mass ratio as $Q = M_{\rm donor}/M_{\rm accretor}$ may differ 
from the definition of the initial mass ratio given by 
$q=M_2/M_1$.}. As we will see, the 
overall effect of the change in the stability criterion on the
population is, however, rather small.

From Table~\ref{param}, it follows that the parameter space occupied
by the bulk of the WDMS binaries and their progenitors is fairly
robust to changes in the mass-accretion parameter $\gamma_{\rm
RLOF}$. Channels~2 and~3 are the most sensitive to the value of the
mass-accretion parameter. The most obvious overall effect of changing
$\gamma_{\rm RLOF}$ is the decrease of the secondary mass $M_2$ at the
birth of the WDMS binaries with increasing degree of
non-conservativeness. In addition, for a given initial primary mass
$M_1$, the increase of the critical mass ratio $Q_{\rm c}$ with
increasing values of $\gamma_{\rm RLOF}$ implies that in models~G05
and~G1 lower initial secondary masses become available for dynamically
stable Roche-lobe overflow. Due to the shape of the initial mass
function [Eq.~(\ref{imf})] and the adopted initial mass ratio
distribution [Eq.~(\ref{imrd})], this behaviour shifts the bulk of
the systems evolving through formation channels~2 and~3 towards lower
primary and secondary masses. An increase in $\gamma_{\rm RLOF}$ is
usually also accompanied by a decrease of the minimum orbital period
at the birth of the WDMS binaries because more angular momentum is
lost from the system. The stellar masses and orbital periods resulting
from fully conservative mass-transfer (model~G0) are furthermore very
close to those found in our standard model (model~A). This concordance
arises from the small values of $\gamma_{\rm RLOF}$ inferred from
Eq.~(\ref{gam}) for binaries in which the thermal time scale of the
accretor is not too much longer than the mass-transfer time scale of
the donor. Finally, we note that the systems evolving through
formation channels~1--3 have typical orbital periods longer than 1
day, in agreement with the orbital period distribution of post-mode II
WDMS binaries derived by de Kool \& Ritter (1993). We recall that the
terminology of mode II mass transfer was introduced by Webbink (1979)
to indicate mass transfer from donor stars with radiative envelopes.

The common-envelope ejection efficiency $\alpha_{\rm CE}$ only affects
the WDMS binaries forming through evolutionary channels~4--6. Since
smaller values of $\alpha_{\rm CE}$ require more orbital energy to
expel the envelope from the system, the minimum orbital period
required for a binary to survive the spiral-in process increases with
decreasing values of $\alpha_{\rm CE}$. For a given donor star mass
$M_1$, the donor star in a binary surviving the common-envelope phase
therefore has more time to evolve when $\alpha_{\rm CE}$ is small, so
that the minimum mass of the nascent white dwarf also increases with
decreasing values of $\alpha_{\rm CE}$. Larger values of $\alpha_{\rm
CE}$, on the other hand, yield wider orbital separations at the end of
the common-envelope phase, making it easier for short-period binaries
to survive the spiral-in process.  These tendencies are clearly seen
in Table~\ref{param} when comparing the variations in the parameter
space between models CE1 ($\alpha_{\rm CE}=0.2$), A ($\alpha_{\rm
CE}=1.0$), and CE8 ($\alpha_{\rm CE}=5.0$).

For a given a given orbital separation $a_{\rm i}$ at the onset of the
common-envelope phase and a given donor star mass $M_1$, the orbital
separation $a_{\rm f}$ at the end of the common-envelope phase
furthermore increases with increasing secondary masses $M_2$. The mass
of the secondary may therefore be the deciding factor determining
whether or not a binary close to the borderline separating merging
from non-merging systems survives the common-envelope phase or
not. From Eq.~(\ref{ce}), it follows that the minimum secondary mass
$M_2$ required to survive the common-envelope phase increases with
decreasing values of $\alpha_{\rm CE}$. This behaviour is responsible
for the variations in the initial secondary mass range of binaries
evolving through formation channel~5. Since $M_1/M_2 > Q_{\rm c}$ is a
necessary requirement for the occurrence of a common-envelope phase
initiated by the primary, the changes in the initial secondary mass
range are accompanied by similar changes in the initial primary mass
range. 

The influence of the donor star's mass $M_1$ on the outcome of the
common-envelope phase can be understood by noting that the radius of
giant-type stars is predominantly determined by the mass of their core, 
and that the radius of the star at the onset of Roche-lobe overflow is
approximately equal to the radius of their Roche lobe. For a given
initial orbital separation $a_{\rm i}$ and a given secondary mass
$M_2$, the orbital separation $a_{\rm f}$ at the end of the
common-envelope phase therefore increases with decreasing values of
$M_1$. Consequently, the maximum primary mass $M_1$ for which a system
may survive the common-envelope phase and the maximum mass of the
resulting white dwarf decrease with decreasing values of $\alpha_{\rm
CE}$. This behaviour is responsible for the lower upper limit on the
initial primary mass range of formation channel~6 when $\alpha_{\rm
CE}=0.2$ (model~CE1).

Besides these general tendencies, the following additional differences
between the various models may be observed upon inspection of the
individual formation channels:

$\bullet$ {\em Channel~2. ---} In model~G0, the minimum initial
primary mass $M_1$ increases from $2\,M_\odot$ to $4\,M_\odot$. The
increase is associated with the destabilising effect of smaller
$\gamma_{\rm RLOF}$-values on the mode of mass transfer from the
primary: since the main-sequence radius of a star increases with
increasing mass, a lower mass star has more time to evolve and thus to
develop a deep convective envelope before it fills its Roche-lobe than
a higher mass star in a binary with the same orbital period. For
$\gamma_{\rm RLOF}=0$, primaries with mass $M_1 \la 4\,M_\odot$
therefore lead to dynamically unstable mass transfer which usually
results in the merger of the primary's core with its main-sequence
companion. We furthermore note that for model~A a gap occurs in the
initial primary mass range between $3\,M_\odot$ and $4\,M_\odot$, in
the white dwarf mass range at the time of formation between
$0.4\,M_\odot$ and $0.7\,M_\odot$, and in the secondary mass range at
the time of formation between $3\,M_\odot$ and $6\,M_\odot$. This
behaviour is caused by the dependency of $\gamma_{\rm RLOF}$ on the
ratio of the mass-transfer time scale of the donor to the thermal time
scale of the accretor [see Eq.~(\ref{gam})]. As a consequence, some
combinations of masses and orbital periods in model~A are subjected to
the same destabilising effect of small $\gamma_{\rm RLOF}$-values as
described for model~G0. Finally, we point that models~A, ~G05 and~G1
yield C/O white dwarfs with masses as low as $\sim
0.3\,M_\odot$. Comparably low-mass C/O white dwarfs were also found by
Iben \& Tutukov (1985) and Han et al. (2000) on the basis of more
detailed numerical calculations. 

$\bullet$ {\em Channel~3. ---} In model~G1, the minimum initial period
of WDMS binary progenitors is significantly shorter than in any of the
other population synthesis models considered. The reason for this is
that in models~A, G0, and~G05, systems with initial orbital periods
shorter than $\sim 600$ days undergo a common-envelope phase instead
of a dynamically stable Roche-lobe overflow phase on the first giant
branch, so that they evolve through formation channel~4 or~5 rather
than through formation channel~3. When mass transfer becomes highly
non-conservative, as in model~G1, the mass-transfer phase becomes
dynamically stable and the primary can avoid losing its entire
envelope until it reaches the AGB.

$\bullet$ {\em Channel~5. ---} In model~G05, the minimum initial
primary mass $M_1$ is about $2\,M_\odot$ lower than in models~G0
and~G1. This behaviour is again associated with the stabilising effect
of larger $\gamma_{\rm RLOF}$-values. Binaries with primary masses
near $4\,M_\odot$ do not survive the common-envelope phase in
model~G0, while they undergo a dynamically stable mass-transfer phase
in model~G1. In the latter case, the systems therefore evolve through
formation channel~2 rather than through formation channel~5. When
$\gamma_{\rm RLOF} \approx 0.5$ (model~G05), mass transfer from
primaries with $M_1 \approx 4\,M_\odot$ initially takes place on the
thermal time scale of the donor star, so that the dynamical
instability is delayed until the donor reaches the red-giant
branch. Contrary to the outcome of the common-envelope phase in
model~G0, the decrease of the donor star's mass during the initial
thermal time scale mass-transfer phase here yields a primary mass at
the onset of the common-envelope phase that allows the binary to
survive the spiral-in process. For a somewhat related reason, the
longest possible initial orbital period decreases from $\sim 1000$
days to $\sim 100$ days in model~CE1. As discussed in
Section~\ref{sch5}, systems with $P_{\rm orb} \la 100$ days start mass
transfer on the thermal time scale of the donor star, while systems
with $P_{\rm orb} \ga 100$ days go straight into the common-envelope
phase. When $\alpha_{\rm CE}=0.2$, the primary masses of the latter
systems at the onset of the common-envelope phase are too high to
avoid a merger during the spiral-in process, so that only the group
with $P_{\rm orb} \la 100$ days survives. Finally, the increase of
the lower limit on the initial orbital periods in model~G1 is
associated with the disappearance of binaries with Hertzsprung-gap
donor stars (i.e. binaries with $P_{\rm orb} \la 200$ days). For
highly non-conservative mass transfer these binaries evolve through
formation channel~2 instead of through formation channel~5.

$\bullet$ {\em Channel~6. ---} In model~G1, the upper limit on the
orbital periods at the formation time of the WDMS binaries decreases with
respect to models~G0 and~G05. The stabilising effect of larger
$\gamma_{\rm RLOF}$-values here implies that less mass needs to be
lost via the stellar wind in order for mass transfer to be dynamically
stable. Many of the longer period systems (i.e. the systems
that have evolved furthest on the AGB and have thus lost the most mass
in a stellar wind) therefore now evolve through formation channel~3
instead of formation channel~6.

\section{WDMS binary numbers and formation rates}
\label{numbers}

Similar to the variations in the parameter space discussed in the
previous section, the assumptions adopted in the various population
synthesis models listed in Table~\ref{models} may affect the formation
rates and the number of WDMS binaries currently populating the
Galaxy. These variations do not apply to systems evolving 
through formation channel~7 since their evolution is governed solely
by mass-loss and mass-accretion from stellar winds. 

\begin{table}
\caption{Total number of WDMS binaries currently populating the Galaxy
  and relative contributions of the interacting and the 
  non-interacting formation channels for different initial mass ratio or
  initial secondary mass distributions. The absolute numbers
  may be converted into an approximate local space density of WDMS
  binaries by dividing them by $5 \times 10^{11}\, {\rm pc^3}$ (see
  Section~\ref{init}).} 
\label{total1}
\begin{tabular}{lccc}
\hline
\hline
      & Total Number    & Interacting     & Non-Interacting \\
Model & (Channel~1--7) & (Channel~1--6) & (Channel~7) \\
\hline
\multicolumn{4}{c}{$n(q) = 1$, $0 < q \le 1$} \\
\hline
A     & 1.6$\times 10^9$ & 13.6\% & 86.4\% \\
G0    & 1.6$\times 10^9$ & 13.4\% & 86.6\% \\
G025  & 1.6$\times 10^9$ & 13.7\% & 86.3\& \\
G05   & 1.6$\times 10^9$ & 14.0\% & 86.0\% \\
G075  & 1.6$\times 10^9$ & 14.6\% & 85.4\% \\
G1    & 1.7$\times 10^9$ & 15.5\% & 84.5\% \\
CE1   & 1.5$\times 10^9$ & \phantom{0}3.5\% & 96.5\% \\
CE3   & 1.6$\times 10^9$ & \phantom{0}9.8\% & 90.2\% \\
CE6   & 1.8$\times 10^9$ & 20.1\% & 79.2\% \\
CE8   & 1.8$\times 10^9$ & 23.4\% & 76.6\% \\
\hline
\hline
\multicolumn{4}{c}{$n(q) \propto q$, $0 < q \le 1$} \\
\hline
A     & 1.3$\times 10^9$ & 15.6\% & 84.4\% \\
G0    & 1.3$\times 10^9$ & 15.4\% & 84.6\% \\
G025  & 1.3$\times 10^9$ & 15.8\% & 84.2\% \\
G05   & 1.3$\times 10^9$ & 16.2\% & 83.8\% \\
G075  & 1.3$\times 10^9$ & 17.2\% & 82.8\% \\
G1    & 1.3$\times 10^9$ & 18.5\% & 81.5\% \\
CE1   & 1.2$\times 10^9$ & \phantom{0}4.9\% & 95.1\% \\
CE3   & 1.2$\times 10^9$ & 11.9\% & 88.1\% \\
CE6   & 1.4$\times 10^9$ & 22.4\% & 77.6\% \\
CE8   & 1.5$\times 10^9$ & 24.7\% & 75.3\% \\
\hline
\hline
\multicolumn{4}{c}{$n(q) \propto q^{-0.99}$, $0 < q \le 1$} \\
\hline
A     & 6.6$\times 10^7$ & 10.7\% & 89.3\% \\
G0    & 6.6$\times 10^7$ & 10.6\% & 89.4\% \\
G025  & 6.6$\times 10^7$ & 10.7\% & 89.3\% \\
G05   & 6.7$\times 10^7$ & 10.9\% & 89.1\% \\
G075  & 6.7$\times 10^7$ & 11.2\% & 88.8\% \\
G1    & 6.7$\times 10^7$ & 11.5\% & 88.5\% \\
CE1   & 6.0$\times 10^7$ & \phantom{0}2.0\% & 98.0\% \\
CE3   & 6.4$\times 10^7$ & \phantom{0}7.0\% & 93.0\% \\
CE6   & 7.3$\times 10^7$ & 18.6\% & 81.4\% \\
CE8   & 7.6$\times 10^7$ & 21.6\% & 78.4\% \\
\hline
\hline
\multicolumn{4}{c}{$M_2$ from IMF given by Eq.~(\ref{imf})} \\
\hline
A     & 2.4$\times 10^9$ & 10.3\% & 89.7\% \\
G0    & 2.4$\times 10^9$ & 10.1\% & 89.9\% \\
G025  & 2.4$\times 10^9$ & 10.3\% & 89.7\% \\
G05   & 2.4$\times 10^9$ & 10.4\% & 89.6\% \\
G075  & 2.4$\times 10^9$ & 10.6\% & 89.4\% \\
G1    & 2.4$\times 10^9$ & 10.7\% & 89.3\% \\
CE1   & 2.2$\times 10^9$ & \phantom{0}1.8\% & 98.2\% \\
CE3   & 2.3$\times 10^9$ & \phantom{0}6.6\% & 93.4\% \\
CE6   & 2.6$\times 10^9$ & 18.2\% & 81.8\% \\
CE8   & 2.7$\times 10^9$ & 21.3\% & 78.7\% \\
\hline
\end{tabular}
\end{table}

The total number of WDMS binaries currently populating the Galaxy and
the relative number of systems that formed through the interacting and
the non-interacting formation channels are listed in
Table~\ref{total1} for different initial mass ratio or initial
secondary mass distributions. The most striking effect is the strong
decrease of the absolute number of systems for the initial mass ratio
distribution $n(q) \propto q^{-0.99}$, for $0 < q \le 1$. The decrease
is related to the larger number of systems undergoing dynamically
unstable mass transfer which ends in the merger of the two component
stars. 

Although the absolute number of systems resulting from formation
channel~7 is independent of the adopted population synthesis model
parameters, the relative number of systems changes due to variations
in the number of systems forming through formation channels~1--6. The
non-interacting systems generally account for 75\% to 85\% of the
total population if the initial mass ratio is distributed according to
$n(q)=1$ or $n(q) \propto q$, for $0 < q \le 1$. This relative number
increases to 95\% when $\alpha_{\rm CE}=0.2$ (model~CE1) due to
the significant decrease of the number of systems surviving the
common-envelope phase in formation channels~4--6. In the cases where
the initial mass ratio is distributed according to $n(q) \propto
q^{-0.99}$, for $0 < q \le 1$, or where the initial secondary mass
$M_2$ is distributed independently from the primary mass $M_1$
according to the initial mass function given by Eq.~(\ref{imf}), the
relative contribution of the non-interacting systems to the population
of WDMS binaries may be even larger.

In Table~\ref{rel1}, the contribution of WDMS binaries with
interacting progenitors to the present-day Galactic population is
further subdivided according to the followed formation channel.  The
absolute number of systems forming through formation channels~1--6
increases with increasing values of $\gamma_{\rm RLOF}$, although the
overall difference is rather small: the number increases by less than
a factor of $\sim 1.25$ between models G0 ($\gamma_{\rm RLOF}=0$) and
G1 ($\gamma_{\rm RLOF}=1$) for all initial mass ratio or initial
secondary mass distributions considered. The effect of $\alpha_{\rm
CE}$ on the absolute number of systems is larger: the number increases
by about an order of magnitude between models CE1 ($\alpha_{\rm
CE}=0.2$) and CE8 ($\alpha_{\rm CE}=5.0$). A similar dependency on
$\alpha_{\rm CE}$ was noted by Iben et al. (1997) for WDMS binaries
with secondaries less massive than $0.3\,M_\odot$.

The relative contributions of formation channels~1--3 increase with
increasing values of $\gamma_{\rm RLOF}$ due to the stabilising effect
of non-conservative mass transfer. Correspondingly, the relative
contributions of channels~4 and~6 decrease with increasing values of
$\gamma_{\rm RLOF}$. The influence of $\gamma_{\rm RLOF}$ is largest
for formation channels~1--3 whose relative contributions increase by
factors of $\sim 5-15$ between models G0 ($\gamma_{\rm RLOF}=0$) and
G1 ($\gamma_{\rm RLOF}=1$).  Increasing the common-envelope ejection
efficiency $\alpha_{\rm CE}$, on the other hand, increases the
relative contributions of channels~4 and~5. The increase is caused by
the larger number of systems surviving the common-envelope phase
resulting from case~B Roche-lobe overflow events.  The relative
contribution of formation channel~6, somewhat surprisingly, decreases
with increasing values of $\alpha_{\rm CE}$, but the absolute number
of systems forming through this channel still increases. The decrease
of the relative number is therefore caused by the more prominent
increase of the absolute number of systems forming through channels~4
and~5. For similar reasons, the relative contributions of
channels~1--3 also decrease with increasing values of $\alpha_{\rm
CE}$.

\begin{table*}
\caption{Total number of present-day Galactic WDMS binaries 
  that formed through formation channels~1--6 and relative contributions
  of each formation channel for different initial mass ratio or
  initial secondary mass distributions. The absolute numbers may be
  converted into an approximate local space density by dividing them
  by $5 \times 10^{11}\, {\rm pc^3}$.}  
\label{rel1}
\begin{tabular}{lccccccc}
\hline
\hline
      &      & \multicolumn{3}{c}{\sc Dynamically stable mass transfer} & \multicolumn{3}{c}{\sc Common\,-\,envelope phase} \\
 & & \multicolumn{2}{c}{\phantom{Ci} $\swarrow$} & $\searrow$ \phantom{Ca C} & \multicolumn{2}{c}{\phantom{Ci} $\swarrow$} & $\searrow$ \phantom{Ca C} \\
 & & \multicolumn{2}{c}{\sc Case B} & {\sc Case C} & \multicolumn{2}{c}{\sc Case B} & {\sc Case C} \\
 & & \phantom{WD} $\swarrow$ & $\searrow$ \phantom{WD} & $\downarrow$ \phantom{i} & \phantom{WD} $\swarrow$ & $\searrow$ \phantom{WD} & $\downarrow$ \phantom{i} \\
 & & {\sc WD} & {\sc nHe star} & {\sc nHe star} & {\sc WD} & {\sc nHe star} & {\sc WD or nHe star} \\
Model & Total Number & (Channel 1) & (Channel 2) & (Channel 3) & (Channel 4) & (Channel 5) & (Channel 6) \\
\hline
\multicolumn{8}{c}{$n(q) = 1$, $0 < q \le 1$} \\
\hline
A     & 2.2$\times 10^8$ & \phantom{1}3.4\% & 0.2\% & 0.1\% & 40.5\% & \phantom{$<$}0.1\% & 55.7\% \\
G0    & 2.2$\times 10^8$ & \phantom{1}2.0\% & 0.2\% & 0.1\% & 41.1\% & \phantom{$<$}0.1\% & 56.5\% \\
G025  & 2.2$\times 10^8$ & \phantom{1}4.1\% & 0.3\% & 0.2\% & 40.1\% & \phantom{$<$}0.2\% & 55.1\% \\
G05   & 2.3$\times 10^8$ & \phantom{1}6.2\% & 0.7\% & 0.4\% & 39.1\% & \phantom{$<$}0.2\% & 53.5\% \\
G075  & 2.4$\times 10^8$ & 10.2\% & 1.6\% & 0.6\% & 37.1\% & \phantom{$<$}0.1\% & 50.5\% \\
G1    & 2.6$\times 10^8$ & 14.8\% & 2.8\% & 1.5\% & 34.4\% & \phantom{$<$}0.1\% & 46.5\% \\
CE1   & 5.2$\times 10^7$ & 14.5\% & 1.0\% & 0.4\% & 18.8\% & $<$0.1\% & 65.2\% \\
CE3   & 1.5$\times 10^8$ & \phantom{1}4.9\% & 0.3\% & 0.1\% & 35.5\% & \phantom{$<$}0.1\% & 59.0\% \\
CE6   & 3.7$\times 10^8$ & \phantom{1}2.0\% & 0.1\% & 0.1\% & 47.1\% & \phantom{$<$}0.4\% & 50.2\% \\
CE8   & 4.3$\times 10^8$ & \phantom{1}1.8\% & 0.1\% & 0.1\% & 49.7\% & \phantom{$<$}0.8\% & 47.6\% \\
\hline
\hline
\multicolumn{8}{c}{$n(q) \propto q$, $0 < q \le 1$} \\
\hline
A     & 2.0$\times 10^8$ & \phantom{1}3.8\% & 0.3\% & 0.2\% & 44.9\% & \phantom{$<$}0.2\% & 50.6\% \\
G0    & 2.0$\times 10^8$ & \phantom{1}2.7\% & 0.3\% & 0.2\% & 45.4\% & \phantom{$<$}0.2\% & 51.2\% \\
G025  & 2.1$\times 10^8$ & \phantom{1}5.2\% & 0.5\% & 0.4\% & 44.2\% & \phantom{$<$}0.2\% & 49.6\% \\
G05   & 2.1$\times 10^8$ & \phantom{1}7.7\% & 1.0\% & 0.6\% & 42.8\% & \phantom{$<$}0.2\% & 47.8\% \\
G075  & 2.3$\times 10^8$ & 12.7\% & 2.2\% & 1.1\% & 39.9\% & \phantom{$<$}0.1\% & 44.1\% \\
G1    & 2.5$\times 10^8$ & 18.6\% & 3.6\% & 2.6\% & 36.0\% & \phantom{$<$}0.1\% & 39.3\% \\
CE1   & 5.7$\times 10^7$ & 13.6\% & 1.2\% & 0.6\% & 20.2\% & $<$0.1\% & 64.3\% \\
CE3   & 1.5$\times 10^8$ & \phantom{1}5.2\% & 0.5\% & 0.2\% & 39.1\% & \phantom{$<$}0.1\% & 54.9\% \\
CE6   & 3.2$\times 10^8$ & \phantom{1}2.4\% & 0.2\% & 0.1\% & 52.6\% & \phantom{$<$}0.5\% & 44.1\% \\
CE8   & 3.6$\times 10^8$ & \phantom{1}2.2\% & 0.2\% & 0.1\% & 55.6\% & \phantom{$<$}0.9\% & 41.1\% \\
\hline
\hline
\multicolumn{8}{c}{$n(q) \propto q^{-0.99}$, $0 < q \le 1$} \\
\hline
A     & 7.1$\times 10^6$ & \phantom{1}2.3\% & 0.1\% & $<$0.1\% & 35.5\% & \phantom{$<$}0.1\% & 62.0\% \\
G0    & 7.0$\times 10^6$ & \phantom{1}1.1\% & 0.1\% & $<$0.1\% & 35.9\% & \phantom{$<$}0.1\% & 62.7\% \\
G025  & 7.1$\times 10^6$ & \phantom{1}2.5\% & 0.2\% & \phantom{$<$}0.1\% & 35.4\% & \phantom{$<$}0.1\% & 61.8\% \\
G05   & 7.2$\times 10^6$ & \phantom{1}3.8\% & 0.4\% & \phantom{$<$}0.1\% & 34.9\% & \phantom{$<$}0.1\% & 60.8\% \\
G075  & 7.5$\times 10^6$ & \phantom{1}6.0\% & 0.8\% & \phantom{$<$}0.2\% & 33.9\% & $<$0.1\% & 59.0\% \\
G1    & 7.7$\times 10^6$ & \phantom{1}8.6\% & 1.6\% & \phantom{$<$}0.6\% & 32.6\% & $<$0.1\% & 56.6\% \\
CE1   & 1.2$\times 10^6$ & 13.3\% & 0.7\% & \phantom{$<$}0.2\% & 17.9\% & $<$0.1\% & 67.9\% \\
CE3   & 4.5$\times 10^6$ & \phantom{1}3.6\% & 0.2\% & \phantom{$<$}0.1\% & 32.0\% & $<$0.1\% & 64.1\% \\
CE6   & 1.4$\times 10^7$ & \phantom{1}1.2\% & 0.1\% & $<$0.1\% & 39.8\% & \phantom{$<$}0.2\% & 58.8\% \\
CE8   & 1.6$\times 10^7$ & \phantom{1}1.0\% & 0.1\% & $<$0.1\% & 41.5\% & \phantom{$<$}0.6\% & 56.9\% \\
\hline
\hline
\multicolumn{8}{c}{$M_2$ from IMF given by Eq.~(\ref{imf})} \\
\hline
A     & 2.5$\times 10^8$ & \phantom{1}2.3\% & $<$0.1\% & $<$0.1\% & 37.0\% & $<$0.1\% & 60.7\% \\
G0    & 2.4$\times 10^8$ & \phantom{1}1.0\% & $<$0.1\% & $<$0.1\% & 37.5\% & $<$0.1\% & 61.5\% \\
G025  & 2.5$\times 10^8$ & \phantom{1}2.3\% & $<$0.1\% & $<$0.1\% & 37.0\% & $<$0.1\% & 60.7\% \\
G05   & 2.5$\times 10^8$ & \phantom{1}3.5\% & \phantom{$<$}0.1\% & \phantom{$<$}0.1\% & 36.5\% & $<$0.1\% & 59.9\% \\
G075  & 2.6$\times 10^8$ & \phantom{1}5.4\% & \phantom{$<$}0.2\% & \phantom{$<$}0.1\% & 35.7\% & $<$0.1\% & 58.6\% \\
G1    & 2.6$\times 10^8$ & \phantom{1}6.9\% & \phantom{$<$}0.4\% & \phantom{$<$}0.3\% & 35.0\% & $<$0.1\% & 57.4\% \\
CE1   & 4.1$\times 10^7$ & 14.0\% & \phantom{$<$}0.1\% & $<$0.1\% & 19.0\% & $<$0.1\% & 66.9\% \\
CE3   & 1.5$\times 10^8$ & \phantom{1}3.7\% & $<$0.1\% & $<$0.1\% & 33.5\% & $<$0.1\% & 62.8\% \\
CE6   & 4.8$\times 10^8$ & \phantom{1}1.2\% & $<$0.1\% & $<$0.1\% & 40.9\% & $<$0.1\% & 57.8\% \\
CE8   & 5.8$\times 10^8$ & \phantom{1}1.0\% & $<$0.1\% & $<$0.1\% & 42.6\% & \phantom{$<$}0.2\% & 56.2\% \\
\hline
\end{tabular}
\end{table*}

The population of WDMS binaries with interacting progenitors is
dominated by systems forming through formation channels~4
and~6. Depending on the adopted model parameters, they generally
account for 80\% to 95\% of the population if the initial mass ratio
is distributed according to $n(q)=1$ or $n(q) \propto q$, for $0 < q
\le 1$, or even for 90\% to 99\% of the population if the initial mass
ratio is distributed according to $n(q) \propto q^{-0.99}$, for $0 < q
\le 1$, or the initial secondary mass is distributed according to the
initial mass function given by Eq.~(\ref{imf}). The short-period
binaries resulting from channels~4 and~6 (see Table~\ref{param})
together with the long-period systems forming from wide
non-interacting progenitors are therefore responsible for the bimodal
nature of the orbital period distribution shown in
Fig.~\ref{1Dfull}. The smallest contributions to the population of
WDMS binaries with interacting progenitors stem from formation 
channels~2, 3 and~5. We note, however, that channels~2 and~5 are
interesting candidates for the formation of WD + early B-star binaries
with long and short orbital periods, respectively
(Fig.~\ref{2DM2}). We will address these channels in this context in a 
forthcoming paper (Willems et al., in preparation).

In Table~\ref{br1}, we list the present-day birthrates of WDMS
binaries forming through each of the considered formation
channels. The birthrate of systems forming through formation channel~7
is independent of the adopted population synthesis model parameters
and amounts to $\sim 0.3\, {\rm yr}^{-1}$, unless the initial
mass ratio is distributed according to $n(q) \propto q^{-0.99}$, for
$0 < q \le 1$. In the latter case, the birthrate of wide
non-interacting systems is of the order of $0.01\, {\rm yr}^{-1}$. The
birthrate derived for our standard model with $n(q)=1$, for $0 < q \le
1$, is in excellent agreement with the birthrate for C/O white dwarfs
in wide binaries of $0.29\, {\rm yr}^{-1}$ derived by Iben \& Tutukov
(1986b). 

\begin{table*}
\caption{Birthrates of Galactic WDMS binaries forming through
  formation channels~1--7 in the case of different initial mass ratio or
  initial secondary mass distributions. The rates may be converted into
  approximate local birthrates by dividing them by $5 \times 10^{11}\,
  {\rm pc^3}$.} 
\label{br1}
\begin{tabular}{lccccccc}
\hline
\hline
 & \multicolumn{3}{c}{\sc Dynamically stable mass transfer} & \multicolumn{3}{c}{\sc Common\,-\,envelope phase} & {\sc Non-}\\
 & \multicolumn{2}{c}{\phantom{Ci} $\swarrow$} & $\searrow$ \phantom{Ca C} & \multicolumn{2}{c}{\phantom{Ci} $\swarrow$} & $\searrow$ \phantom{Ca C} & \\
 & \multicolumn{2}{c}{\sc Case B} & {\sc Case C} & \multicolumn{2}{c}{\sc Case B} & {\sc Case C} & {\sc Interacting} \\
 & \phantom{WD} $\swarrow$ & $\searrow$ \phantom{WD} & $\downarrow$ \phantom{i} & \phantom{WD} $\swarrow$ & $\searrow$ \phantom{WD} & $\downarrow$ \phantom{i} & \\
 & {\sc WD} & {\sc nHe star} & {\sc nHe star} & {\sc WD} & {\sc nHe star} & {\sc WD or nHe star} & {\sc Systems} \\
Model & (Channel 1) & (Channel 2) & (Channel 3) & (Channel 4) & (Channel 5) & (Channel 6) & (Channel 7)\\
\hline
\multicolumn{8}{c}{$n(q) = 1$, $0 < q \le 1$} \\
\hline
A     & 5.0$\times 10^{-3}$ & 1.0$\times 10^{-3}$ & 3.5$\times 10^{-4}$ & 3.5$\times 10^{-2}$ & 5.2$\times 10^{-4}$ & 3.4$\times 10^{-2}$ & 3.0$\times 10^{-1}$ \\
G0    & 4.6$\times 10^{-3}$ & 9.6$\times 10^{-4}$ & 3.5$\times 10^{-4}$ & 3.5$\times 10^{-2}$ & 5.1$\times 10^{-4}$ & 3.4$\times 10^{-2}$ & 3.0$\times 10^{-1}$ \\
G025  & 6.8$\times 10^{-3}$ & 1.5$\times 10^{-3}$ & 5.9$\times 10^{-4}$ & 3.5$\times 10^{-2}$ & 6.5$\times 10^{-4}$ & 3.4$\times 10^{-2}$ & 3.0$\times 10^{-1}$ \\
G05   & 8.0$\times 10^{-3}$ & 3.1$\times 10^{-3}$ & 8.8$\times 10^{-4}$ & 3.5$\times 10^{-2}$ & 7.4$\times 10^{-4}$ & 3.4$\times 10^{-2}$ & 3.0$\times 10^{-1}$ \\
G075  & 1.0$\times 10^{-2}$ & 6.2$\times 10^{-3}$ & 1.3$\times 10^{-3}$ & 3.5$\times 10^{-2}$ & 2.5$\times 10^{-4}$ & 3.3$\times 10^{-2}$ & 3.0$\times 10^{-1}$ \\
G1    & 1.2$\times 10^{-2}$ & 7.5$\times 10^{-3}$ & 2.7$\times 10^{-3}$ & 3.5$\times 10^{-2}$ & 2.6$\times 10^{-4}$ & 3.2$\times 10^{-2}$ & 3.0$\times 10^{-1}$ \\
CE1   & 5.0$\times 10^{-3}$ & 1.0$\times 10^{-3}$ & 3.5$\times 10^{-4}$ & 8.3$\times 10^{-3}$ & 2.9$\times 10^{-6}$ & 1.3$\times 10^{-2}$ & 3.0$\times 10^{-1}$ \\
CE3   & 5.0$\times 10^{-3}$ & 1.0$\times 10^{-3}$ & 3.5$\times 10^{-4}$ & 2.6$\times 10^{-2}$ & 1.8$\times 10^{-4}$ & 2.7$\times 10^{-2}$ & 3.0$\times 10^{-1}$ \\
CE6   & 5.0$\times 10^{-3}$ & 1.0$\times 10^{-3}$ & 3.5$\times 10^{-4}$ & 5.6$\times 10^{-2}$ & 2.4$\times 10^{-3}$ & 4.4$\times 10^{-2}$ & 3.0$\times 10^{-1}$ \\
CE8   & 5.0$\times 10^{-3}$ & 1.0$\times 10^{-3}$ & 3.5$\times 10^{-4}$ & 6.5$\times 10^{-2}$ & 4.3$\times 10^{-3}$ & 4.4$\times 10^{-2}$ & 3.0$\times 10^{-1}$ \\
\hline
\hline
\multicolumn{8}{c}{$n(q) \propto q$, $0 < q \le 1$} \\
\hline
A     & 6.6$\times 10^{-3}$ & 1.4$\times 10^{-3}$ & 5.6$\times 10^{-4}$ & 3.8$\times 10^{-2}$ & 6.6$\times 10^{-4}$ & 3.5$\times 10^{-2}$ & 3.0$\times 10^{-1}$ \\
G0    & 6.3$\times 10^{-3}$ & 1.3$\times 10^{-3}$ & 5.6$\times 10^{-4}$ & 3.8$\times 10^{-2}$ & 6.4$\times 10^{-4}$ & 3.5$\times 10^{-2}$ & 3.0$\times 10^{-1}$ \\
G025  & 9.1$\times 10^{-3}$ & 2.0$\times 10^{-3}$ & 9.8$\times 10^{-4}$ & 3.8$\times 10^{-2}$ & 8.0$\times 10^{-4}$ & 3.5$\times 10^{-2}$ & 3.0$\times 10^{-1}$ \\
G05   & 1.1$\times 10^{-2}$ & 4.0$\times 10^{-3}$ & 1.4$\times 10^{-3}$ & 3.8$\times 10^{-2}$ & 9.1$\times 10^{-4}$ & 3.4$\times 10^{-2}$ & 3.0$\times 10^{-1}$ \\
G075  & 1.4$\times 10^{-2}$ & 8.3$\times 10^{-3}$ & 2.2$\times 10^{-3}$ & 3.8$\times 10^{-2}$ & 2.9$\times 10^{-4}$ & 3.3$\times 10^{-2}$ & 3.0$\times 10^{-1}$ \\
G1    & 1.5$\times 10^{-2}$ & 1.0$\times 10^{-2}$ & 4.6$\times 10^{-3}$ & 3.7$\times 10^{-2}$ & 2.9$\times 10^{-4}$ & 3.2$\times 10^{-2}$ & 3.0$\times 10^{-1}$ \\
CE1   & 6.6$\times 10^{-3}$ & 1.4$\times 10^{-3}$ & 5.6$\times 10^{-4}$ & 9.4$\times 10^{-3}$ & 3.2$\times 10^{-6}$ & 1.5$\times 10^{-2}$ & 3.0$\times 10^{-1}$ \\
CE3   & 6.6$\times 10^{-3}$ & 1.4$\times 10^{-3}$ & 5.6$\times 10^{-4}$ & 2.8$\times 10^{-2}$ & 2.3$\times 10^{-4}$ & 2.9$\times 10^{-2}$ & 3.0$\times 10^{-1}$ \\
CE6   & 6.6$\times 10^{-3}$ & 1.4$\times 10^{-3}$ & 5.6$\times 10^{-4}$ & 5.9$\times 10^{-2}$ & 2.8$\times 10^{-3}$ & 4.4$\times 10^{-2}$ & 3.0$\times 10^{-1}$ \\
CE8   & 6.6$\times 10^{-3}$ & 1.4$\times 10^{-3}$ & 5.6$\times 10^{-4}$ & 6.8$\times 10^{-2}$ & 4.7$\times 10^{-3}$ & 4.4$\times 10^{-2}$ & 3.0$\times 10^{-1}$ \\
\hline
\hline
\multicolumn{8}{c}{$n(q) \propto q^{-0.99}$, $0 < q \le 1$} \\
\hline
A     & 8.3$\times 10^{-5}$ & 1.5$\times 10^{-5}$ & 4.5$\times 10^{-6}$ & 9.2$\times 10^{-4}$ & 8.8$\times 10^{-6}$ & 1.0$\times 10^{-3}$ & 9.8$\times 10^{-3}$ \\
G0    & 7.3$\times 10^{-5}$ & 1.4$\times 10^{-5}$ & 4.5$\times 10^{-6}$ & 9.2$\times 10^{-4}$ & 8.5$\times 10^{-6}$ & 1.0$\times 10^{-3}$ & 9.8$\times 10^{-3}$ \\
G025  & 1.1$\times 10^{-4}$ & 2.2$\times 10^{-5}$ & 7.4$\times 10^{-6}$ & 9.2$\times 10^{-4}$ & 1.1$\times 10^{-5}$ & 1.0$\times 10^{-3}$ & 9.8$\times 10^{-3}$ \\
G05   & 1.3$\times 10^{-4}$ & 4.9$\times 10^{-5}$ & 1.1$\times 10^{-5}$ & 9.2$\times 10^{-4}$ & 1.3$\times 10^{-5}$ & 1.0$\times 10^{-3}$ & 9.8$\times 10^{-3}$ \\
G075  & 1.7$\times 10^{-4}$ & 9.5$\times 10^{-5}$ & 1.6$\times 10^{-5}$ & 9.2$\times 10^{-4}$ & 4.9$\times 10^{-6}$ & 1.0$\times 10^{-3}$ & 9.8$\times 10^{-3}$ \\
G1    & 1.8$\times 10^{-4}$ & 1.2$\times 10^{-4}$ & 3.3$\times 10^{-5}$ & 9.2$\times 10^{-4}$ & 5.0$\times 10^{-6}$ & 9.9$\times 10^{-4}$ & 9.8$\times 10^{-3}$ \\
CE1   & 8.3$\times 10^{-5}$ & 1.5$\times 10^{-5}$ & 4.5$\times 10^{-6}$ & 1.8$\times 10^{-4}$ & 5.0$\times 10^{-8}$ & 3.3$\times 10^{-4}$ & 9.8$\times 10^{-3}$ \\
CE3   & 8.3$\times 10^{-5}$ & 1.5$\times 10^{-5}$ & 4.5$\times 10^{-6}$ & 6.5$\times 10^{-4}$ & 2.9$\times 10^{-6}$ & 7.8$\times 10^{-4}$ & 9.8$\times 10^{-3}$ \\
CE6   & 8.3$\times 10^{-5}$ & 1.5$\times 10^{-5}$ & 4.5$\times 10^{-6}$ & 1.5$\times 10^{-3}$ & 4.5$\times 10^{-5}$ & 1.4$\times 10^{-3}$ & 9.8$\times 10^{-3}$ \\
CE8   & 8.3$\times 10^{-5}$ & 1.5$\times 10^{-5}$ & 4.5$\times 10^{-6}$ & 1.8$\times 10^{-3}$ & 9.0$\times 10^{-5}$ & 1.5$\times 10^{-3}$ & 9.8$\times 10^{-3}$ \\
\hline
\hline
\multicolumn{8}{c}{$M_2$ from IMF given by Eq.~(\ref{imf})} \\
\hline
A     & $2.3\times 10^{-3}$ & $4.0\times 10^{-5}$ & $2.0\times 10^{-5}$ & $3.3\times 10^{-2}$ & $1.7\times 10^{-5}$ & $3.1\times 10^{-2}$ & 3.3$\times 10^{-1}$ \\
G0    & $1.9\times 10^{-3}$ & $2.7\times 10^{-5}$ & $2.0\times 10^{-5}$ & $3.3\times 10^{-2}$ & $1.5\times 10^{-5}$ & $3.1\times 10^{-2}$ & 3.3$\times 10^{-1}$ \\
G025  & $3.0\times 10^{-3}$ & $5.8\times 10^{-5}$ & $6.0\times 10^{-5}$ & $3.3\times 10^{-2}$ & $2.7\times 10^{-5}$ & $3.1\times 10^{-2}$ & 3.3$\times 10^{-1}$ \\
G05   & $3.3\times 10^{-3}$ & $2.4\times 10^{-4}$ & $1.2\times 10^{-4}$ & $3.3\times 10^{-2}$ & $3.4\times 10^{-5}$ & $3.1\times 10^{-2}$ & 3.3$\times 10^{-1}$ \\
G075  & $4.1\times 10^{-3}$ & $5.2\times 10^{-4}$ & $2.0\times 10^{-4}$ & $3.3\times 10^{-2}$ & $1.1\times 10^{-5}$ & $3.0\times 10^{-2}$ & 3.3$\times 10^{-1}$ \\
G1    & $4.3\times 10^{-3}$ & $6.7\times 10^{-4}$ & $5.2\times 10^{-4}$ & $3.3\times 10^{-2}$ & $1.2\times 10^{-5}$ & $3.0\times 10^{-2}$ & 3.3$\times 10^{-1}$ \\
CE1   & $2.3\times 10^{-3}$ & $4.0\times 10^{-5}$ & $2.0\times 10^{-5}$ & $6.7\times 10^{-3}$ & $6.2\times 10^{-8}$ & $1.0\times 10^{-2}$ & 3.3$\times 10^{-1}$ \\
CE3   & $2.3\times 10^{-3}$ & $4.0\times 10^{-5}$ & $2.0\times 10^{-5}$ & $2.3\times 10^{-2}$ & $4.5\times 10^{-6}$ & $2.4\times 10^{-2}$ & 3.3$\times 10^{-1}$ \\
CE6   & $2.3\times 10^{-3}$ & $4.0\times 10^{-5}$ & $2.0\times 10^{-5}$ & $5.5\times 10^{-2}$ & $2.0\times 10^{-4}$ & $4.2\times 10^{-2}$ & 3.3$\times 10^{-1}$ \\
CE8   & $2.3\times 10^{-3}$ & $4.0\times 10^{-5}$ & $2.0\times 10^{-5}$ & $6.5\times 10^{-2}$ & $7.7\times 10^{-4}$ & $4.5\times 10^{-2}$ & 3.3$\times 10^{-1}$ \\
\hline
\end{tabular}
\end{table*}

The general dependencies of the birthrates of formation channels~1--6
on the adopted population synthesis model are similar to those of the
absolute and relative numbers of systems described above. The effect
of changing $\gamma_{\rm RLOF}$ is largest for channels~2 and~3 where
the birthrate increases by an order of magnitude between models~G0
($\gamma_{\rm RLOF}=0$) and G1 ($\gamma_{\rm RLOF}=1$). The impact of
$\gamma_{\rm RLOF}$ on the other channels is typically smaller than a
factor of $\sim 3$. Changes in $\alpha_{\rm CE}$, on the other hand,
predominantly affect formation channels~4 and~5. The birthrates of
systems forming through these channels increase by 1 and 3 orders of
magnitude, respectively, between model CE1 ($\alpha_{\rm CE}=0.2$) and
model CE8 ($\alpha_{\rm CE}=5.0$). In the case where the initial
secondary mass is distributed according to the same initial mass
function as the initial primary mass, the birthrate of channel~5
increases by 4 orders of magnitude. The birthrate of systems forming
through formation channel~6 varies by less than a factor of $\sim 5$,
while the birthrates of systems forming through formation
channels~1--3 are not affected at all. For $\gamma_{\rm RLOF} \la
0.25$ and $\alpha_{\rm CE} \ga 0.6$, the total birthrate of all WDMS
binaries forming through a common-envelope phase (channels~4--6) is
furthermore about 10 times larger than the total birthrate of all WDMS
binaries forming through a stable Roche-lobe overflow phase
(channels~1--3). This is again in excellent agreement with de Kool \&
Ritter (1993).

\begin{figure*}
\resizebox{8.4cm}{!}{\includegraphics{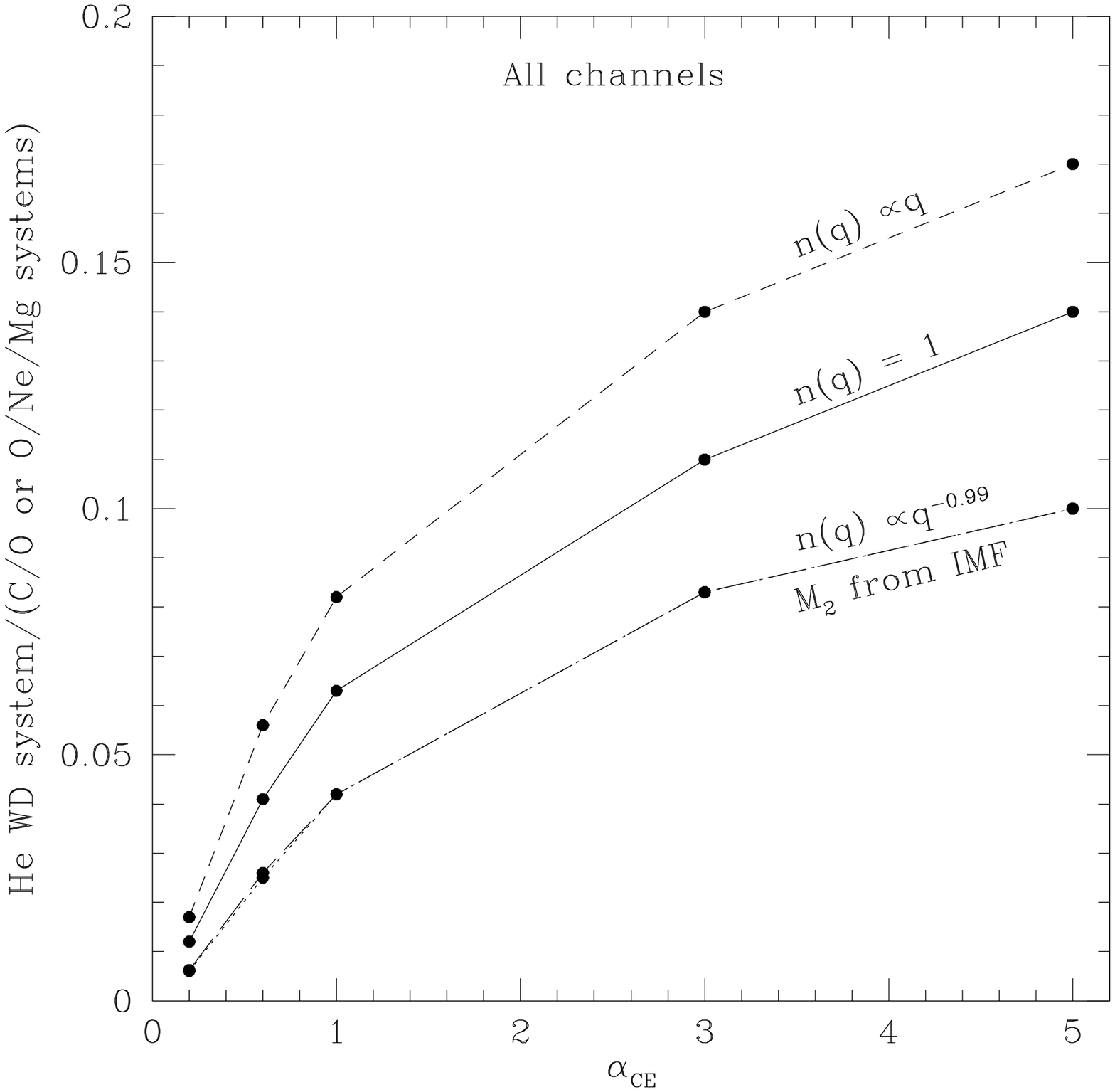}}
\resizebox{8.4cm}{!}{\includegraphics{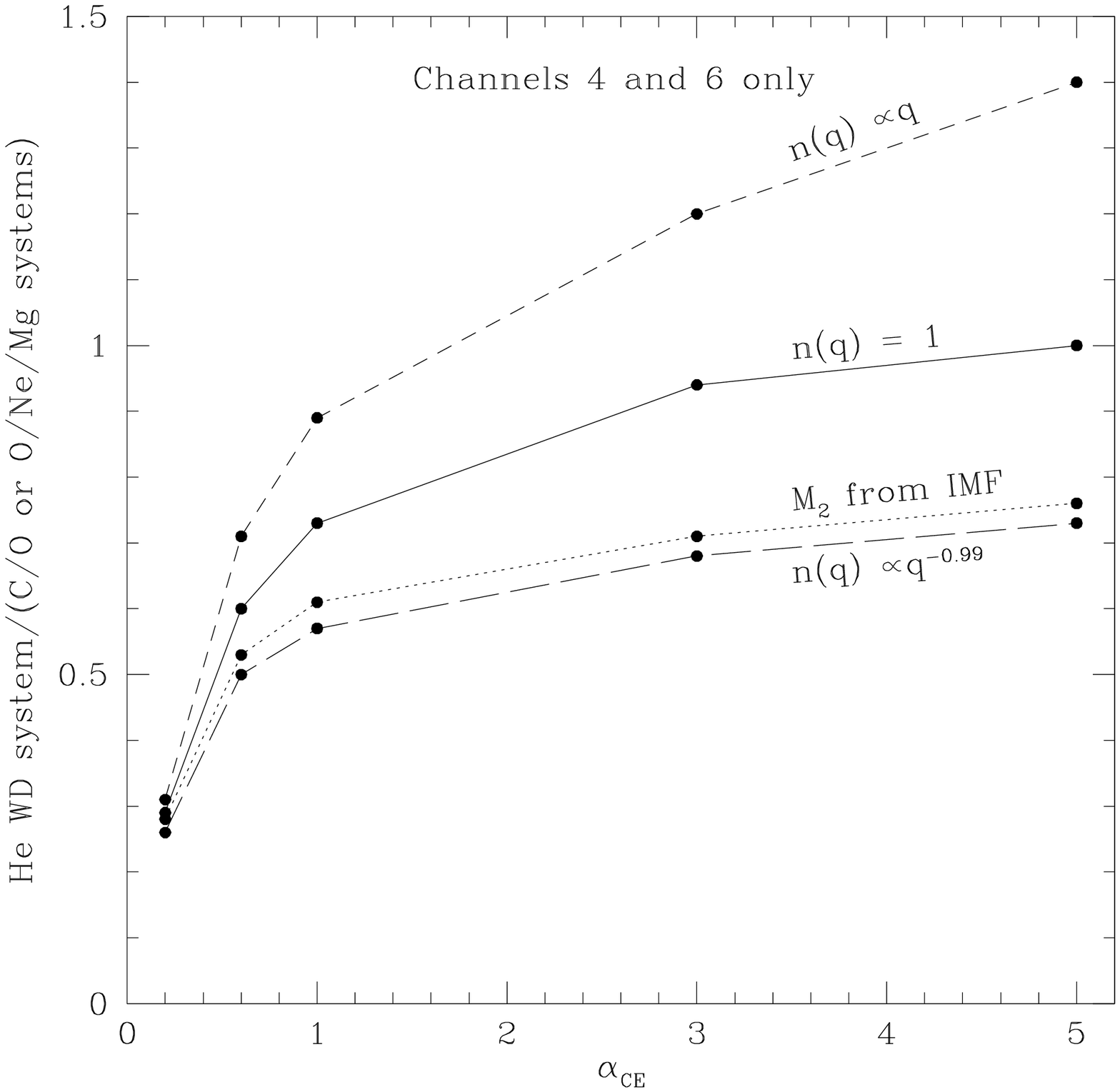}}
\caption{Ratio of He white dwarf WDMS binaries to C/O or O/Ne/Mg WDMS
  binaries for different initial mass ratio or initial secondary mass
  distributions. The left-hand panel shows the ratios as a function of
  the common-envelope ejection efficiency $\alpha_{\rm CE}$ for WDMS
  binaries forming through all considered formation channels, while
  the right-hand panel shows the ratios for systems forming through
  formation channels~4 and~6 only. In the left-hand panel, the
  curves associated with $n(q) \propto q^{-0.99}$ and with $M_2$ drawn
  independently from the same IMF as $M_1$ are almost
  indistinguishable.}
\label{fratio}
\end{figure*}

For conclusion, we consider the ratio of the number of systems
containing a He white dwarf to the number of systems containing a C/O
or O/Ne/Mg white dwarf. This ratio is independent of the normalisation
of the adopted star formation rate [Eq.~(\ref{sfr})] and of the
fraction of stars in binaries. Due to the dominance of the wide
non-interacting systems and the systems forming through the
common-envelope channels~4 and~6, the ratio is fairly insensitive to
changes in the mass-accretion parameter $\gamma_{\rm RLOF}$. The role
of the common-envelope ejection parameter $\alpha_{\rm CE}$, on the
other hand, is much more prominent. This is illustrated in the
left-hand panel of Fig.~\ref{fratio}. For $\alpha_{\rm CE}=0.2$ (model
CE1) the ratio takes comparably small values for each of the initial
mass ratio or initial secondary mass distributions
considered. Provided that a complete observational sample of WDMS
binaries can be compiled and the nature of the white dwarf can be
established, comparison of the observationally derived ratio of He
white dwarf systems to C/O or O/Ne/Mg white dwarf systems with the
theoretically predicted values may therefore constrain the
common-envelope ejection efficiency. For higher values of $\alpha_{\rm
CE}$, the effect of changing $\alpha_{\rm CE}$ can not be
distinguished from the effect of changing the adopted initial mass
ratio or initial secondary mass distribution.  As shown in the
right-hand panel of Fig.~\ref{fratio} similar dependency on
$\gamma_{\rm RLOF}$ and $\alpha_{\rm CE}$ is also found when the
subset of short-period systems resulting from the dominant channels~4
and~6 is considered. This subset is potentially more interesting than
the whole sample of WDMS binaries since shorter period systems are
easier to detect observationally than longer period systems.

\section{Luminosity ratios and radial velocities}

The detection and identification of WDMS binaries generally relies on
the ability to detect the white dwarf's signature in the composite
binary spectrum and to determine the components' orbital
radial-velocity variations. Regardless of any additional selection
effects, the detection of the white dwarf's spectral signature becomes
increasingly difficult with decreasing ratio of the white dwarf's
luminosity to the main-sequence star's luminosity, while the detection
of radial-velocity variations becomes more difficult with increasing
orbital periods.

\begin{figure*}
\begin{center}
\resizebox{8.4cm}{!}{\includegraphics{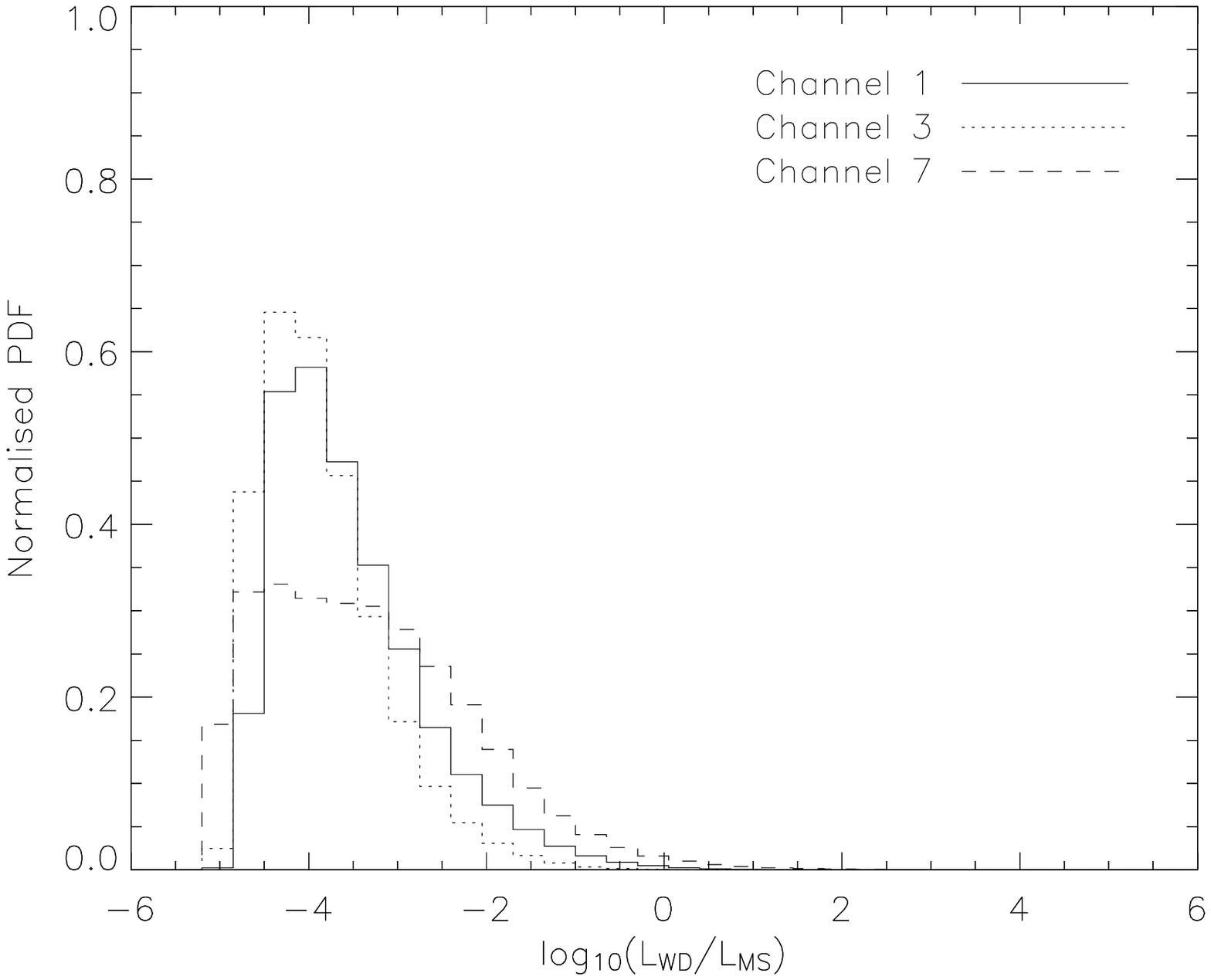}}
\resizebox{8.4cm}{!}{\includegraphics{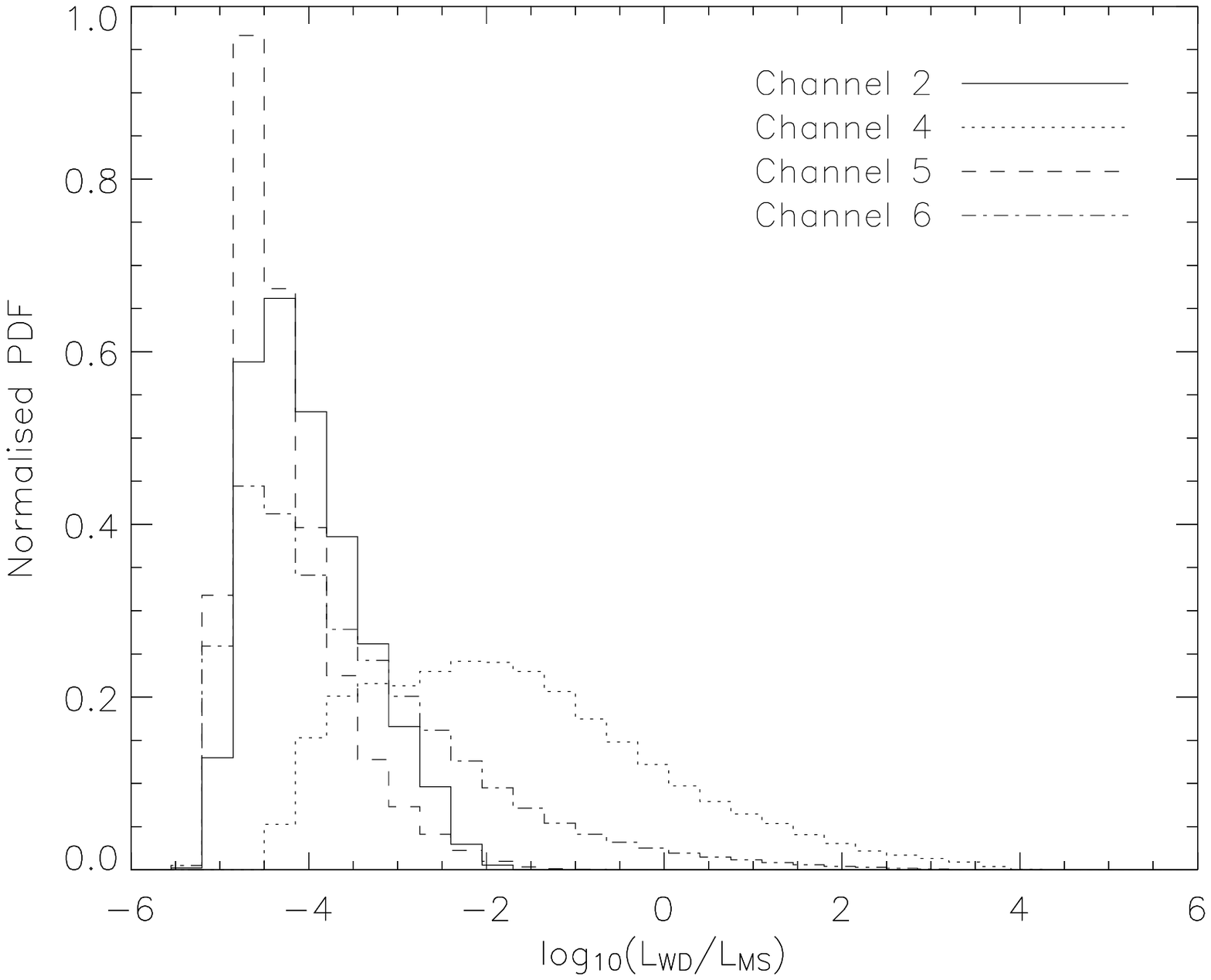}} \\
\caption{Normalised distribution functions of white dwarf
  to main-sequence star luminosity ratios for formation channels 1--7
  in the case of population synthesis model~A and the initial mass
  ratio distribution $n(q)=1$.}
\label{LWDLMSpdf}
\end{center}
\end{figure*}

The luminosity ratio may be particularly stringent for binaries with
older white dwarfs since the time during which a white dwarf is
detectable is typically of the order of $10^8$ years or shorter
(e.g. Iben et al. 1997). The distribution of the luminosity ratios
$L_{\rm WD}/L_{\rm MS}$, where $L_{\rm WD}$ is the luminosity of the
white dwarf and $L_{\rm MS}$ that of its main-sequence companion, for
WDMS binaries forming through the different formation channels are
displayed in Fig.~\ref{LWDLMSpdf} in the case of population synthesis
model~A and the initial mass ratio distribution $n(q)=1$.  For the
determination of the luminosity ratios, we adopted the expression for
white dwarf cooling by Hurley et al. (2000) and neglected the
relatively small increase of the secondary's luminosity during its
evolution on the main sequence. The luminosity ratio distributions
typically peak between $10^{-5}$ and $10^{-3}$, except in the case of
formation channel~4 where a much broader distribution is found which
extends up to significantly higher luminosity ratios. The figure
furthermore shows that channels~4 and ~6 contain a significant number
of WDMS binaries with small luminosity ratios, so that it may actually
be hard to compile a complete sample of systems to constrain the 
common-envelope ejection efficiency $\alpha_{\rm CE}$.

\begin{figure*}
\begin{center}
\resizebox{8.4cm}{!}{\includegraphics{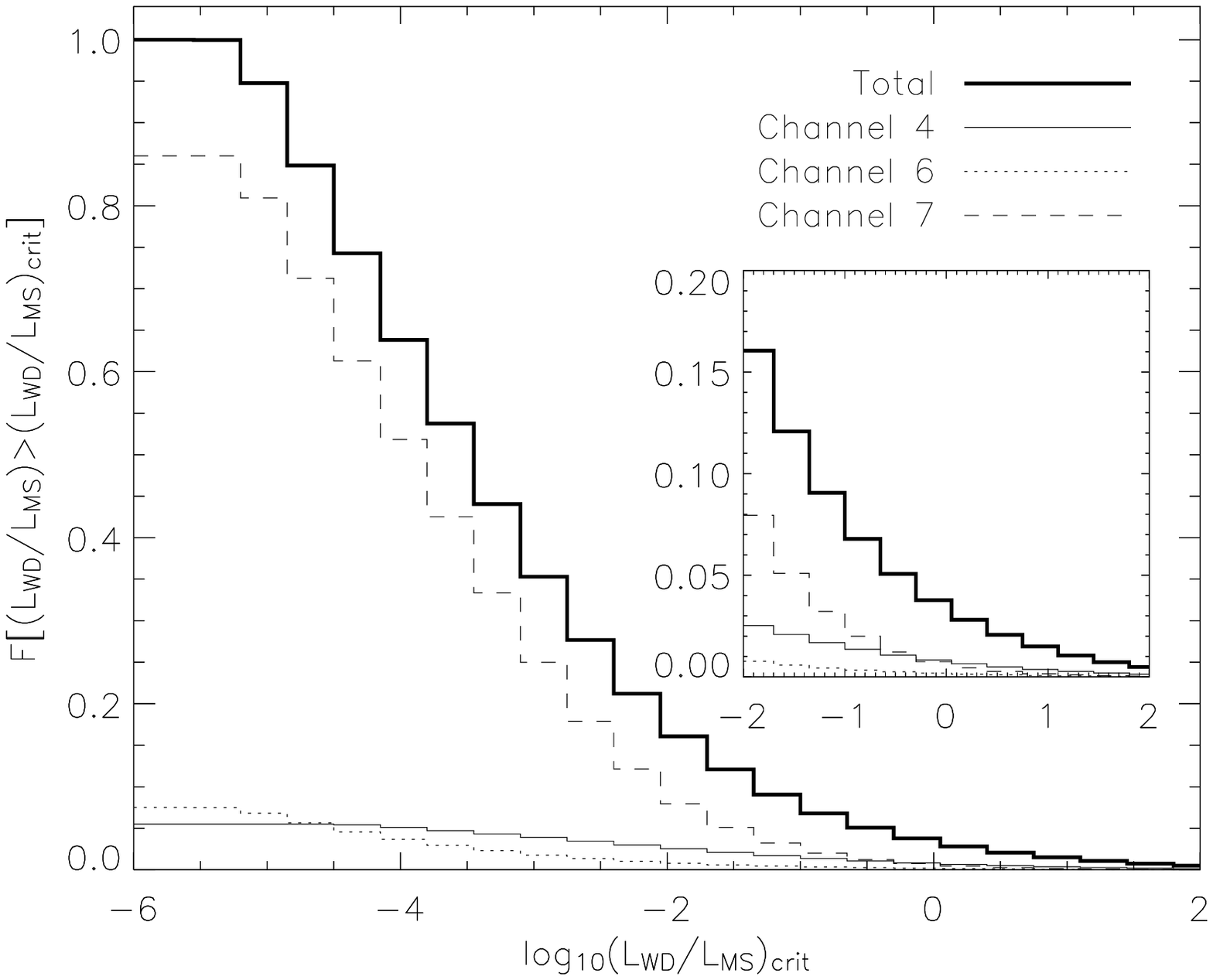}}
\resizebox{8.4cm}{!}{\includegraphics{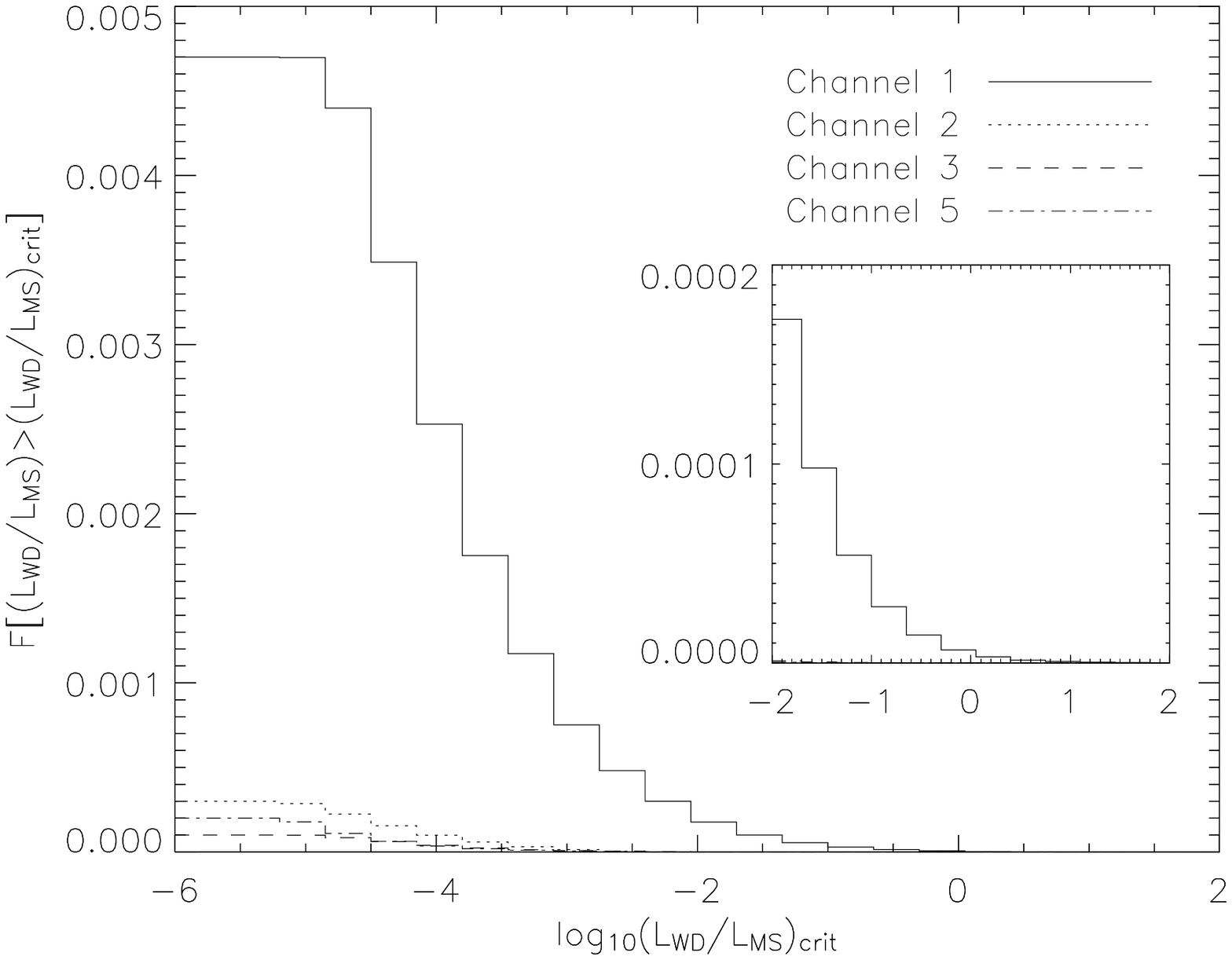}} \\
\caption{The fraction $F$ of systems for wich the white dwarf to
  main-sequence star luminosity ratio is larger than the limiting
  ratio $\left( L_{\rm WD}/L_{\rm MS} \right)_{\rm crit}$, in the case
  of population synthesis model~A and the initial mass ratio
  distribution $n(q)=1$. The thick solid line represents the total for
  all systems forming through formation channels 1--7, while the thin
  lines distinguish between the contributions of the different
  formation channels. The inserts zoom in on the regions where the
  distribution functions are small. } 
\label{LWDLMScr}
\end{center}
\end{figure*}

In Fig.~\ref{LWDLMScr}, we combine the luminosity ratio distributions
shown in Fig.~\ref{LWDLMSpdf} with the relative contributions of the
different formation channels to the WDMS binary population given in
Tables~\ref{total1} and~\ref{rel1} to determine the fraction of
systems for wich the white dwarf to main-sequence star luminosity
ratio is larger than a given limiting ratio $\left( L_{\rm WD}/L_{\rm
MS} \right)_{\rm crit}$. The thick solid line represents the total for
all systems forming through formation channels 1--7, while the thin
lines distinguish between the contributions of the individual
formation channels. For $\left( L_{\rm WD}/L_{\rm MS} \right)_{\rm
crit}=10^{-6}$, the fractional contributions of the different channels
correspond to those obtained from Tables~\ref{total1}
and~\ref{rel1}.  The wide non-interacting systems dominate up to
$\left( L_{\rm WD}/L_{\rm MS} \right)_{\rm crit} \approx 10^{-1}$.
For larger values of $\left( L_{\rm WD}/L_{\rm MS} \right)_{\rm
crit}$, the systems forming through formation channel~4 become equally
important. 

The amplitudes $K_{\rm WD}$ and $K_{\rm MS}$ of the white dwarf and
the main-sequence star's orbital radial-velocity variations in the
case of a typical orbital inclination $i=60^\circ$ and under the
assumption of a fixed Keplerian orbit are listed in
Table~\ref{KL}. This assumption breaks down for short-period systems
in which magnetic braking and/or gravitational radiation cause the
components to spiral-in towards each other.  However, we do not expect
this to significantly affect our radial-velocity estimates, unless the
orbital evolution time scale becomes shorter than or comparable to the
white dwarf cooling time. As present-day radial-velocity surveys
easily reach accuracies down to 10 ${\rm km\, s^{-1}}$, all formation
channels except channels~3 and~7 produce WDMS binaries with readily
detectable radial-velocity amplitudes. If the spectra of the component
stars can be disentangled, the common-envelope scenarios furthermore
produce WDMS binaries in which both components may have observable
radial-velocity variations.

\begin{table*}
\caption{Typical orbital radial-velocity amplitudes of WDMS binary
  components in the case of population synthesis model~A and the
  initial mass ratio distribution $n(q)=1$. The radial-velocity
  amplitudes correspond to a typical orbital inclination of
  $60^\circ$.}  
\label{KL}
\begin{tabular}{lrr}
\hline
\hline
Formation channel & $K_{\rm WD}$\phantom{000} & $K_{\rm MS}$\phantom{000} \\
\hline
1 (Stable case B RLOF $\rightarrow$ WD) & $80-100$ ${\rm km\, s^{-1}}$ & $5-10$ ${\rm km\, s^{-1}}$ \\
2 (Stable case B RLOF $\rightarrow$ nHe star) & $50-100$ ${\rm km\, s^{-1}}$ & $5-10$ ${\rm km\, s^{-1}}$ \\
3 (Stable case C RLOF $\rightarrow$ nHe star) & $5-15$ ${\rm km\, s^{-1}}$   & $2-5$ ${\rm km\, s^{-1}}$ \\
4 (Case B CE $\rightarrow$ WD) & $70-220$ ${\rm km\, s^{-1}}$ & $30-140$ ${\rm km\, s^{-1}}$ \\
5 (Case B CE $\rightarrow$ nHe star) &     $170-220$ ${\rm km\, s^{-1}}$ &     $20-30$ ${\rm km\, s^{-1}}$ \\
6 (Case C CE $\rightarrow$ nHe star or WD) &  $40-170$ ${\rm km\, s^{-1}}$ & $20-140$ ${\rm km\, s^{-1}}$ \\
7 (Non-interacting systems) &     $1-10$ ${\rm km\, s^{-1}}$ &   $1-5$ ${\rm km\, s^{-1}}$ \\
\hline
\end{tabular}
\end{table*}

\section{Concluding remarks}

We used the BiSEPS binary population synthesis code described by
Willems \& Kolb (2002) to study the population of detached white dwarf
main-sequence star binaries forming through seven distinct
evolutionary channels. In six of these, the white dwarf is formed
through binary interactions resulting from Roche-lobe overflow. The
channels mainly differ in the stability and the outcome of the
mass-transfer phase that eventually gives rise to the formation of the
white dwarf. The six formation channels are depicted schematically in
Figs.~\ref{ch1} and \ref{ch2}--\ref{ch6}. The seventh evolutionary
channel is characterised by the absence of any type of binary
interactions other than mass and angular momentum exchange via a
stellar wind. Binaries evolving through this channel are typically
very wide, so that the white dwarf forms in much the same way as it
would if the primary were a single star.

Our results show that the wide non-interacting binaries generally
comprise more than 75\% of the total WDMS binary population. The
remaining part of the population is dominated by systems undergoing a
common-envelope phase when the white dwarf progenitor ascends the
first giant branch or the asymptotic giant branch. The total number of
systems currently populating the Galaxy, the relative number of
systems evolving through each formation channel, and the birthrates of
systems forming through each channel are given in
Tables~\ref{total1}--\ref{br1} for different different assumptions
about the fate of the mass transferred during dynamically stable
Roche-lobe overflow, different common-envelope ejection efficiencies,
and different initial mass ratio or secondary mass distributions. The
total number of systems is of the order of $\sim 10^9$ when a flat
initial mass ratio distribution $n(q) = 1$, for $0 < q \le 1$, is
considered. This number decreases to $\sim 7 \times 10^7$ for an
initial mass ratio distribution of the form $n(q) \propto q^{-0.99}$,
for $0 < q \le 1$. An even stronger decrease is found for the
birthrates of systems forming through the different formation
channels. Besides the different assumptions regarding the treatment of
dynamically stable and unstable mass transfer, the total number of
WDMS binaries currently populating the Galaxy also depends on the
finite life time of the systems which, in the case of short-period
systems, may be affected by the adopted magnetic braking
prescription. A detailed analysis of the effects of different braking
laws is, however, beyond the scope of this investigation.

The total number of Galactic WDMS binaries evolving through a
common-envelope phase is of the order of $\sim 2 \times 10^8$
when a flat initial mass ratio distribution $n(q) = 1$ is considered,
and of the order of $\sim 7 \times 10^6$ when an initial mass ratio
distribution of the form $n(q) \propto q^{-0.99}$ is considered. Under
the assumption that the Galaxy has an effective volume of $5 \times
10^{11}\, {\rm pc^3}$, these numbers correspond to approximate local
space densities of $4 \times 10^{-4}\, {\rm pc^{-3}}$ and
$10^{-5}\, {\rm pc^{-3}}$, respectively. The latter space density is
comparable to the space density of $6 - 30 \times 10^{-6}\, {\rm
pc^{-3}}$ derived by Schreiber \& G\"ansicke (2003) from an observed
sample of 30 post-common-envelope binaries with low-mass main-sequence
secondaries. However, since the space densities derived by Schreiber
\& G\"ansicke (2003) are likely to be lower limits, the space density
of $4 \times 10^{-4}\, {\rm pc^{-3}}$ is not necessarily in
disagreement with observations. We also note that 12 of the 30
post-common-envelope binaries considered by Schreiber \& G\"ansicke
(2003) contain an sdOB primary instead of a white dwarf. These sdOB +
main-sequence star systems are not included in our simulated WDMS
binary samples.

We furthermore find that the ratio of the number of He white dwarf
systems to the number of C/O or O/Ne/Mg white dwarf systems is weakly
dependent on the amount of mass lost from the system during
dynamically stable Roche-lobe overflow and strongly dependent on the
common-envelope ejection efficiency. We therefore propose that
comparison of the observationally derived ratio from, e.g.,
large-scale planet-search mission as SuperWASP, COROT, and Kepler with
the ratios presented in Fig.~\ref{fratio} may yield constraints on the
common-envelope ejection efficiency, provided that the efficiency is
low. For higher efficiencies, the role of the ejection efficiency
parameter cannot be distinguished from the role of the initial mass
ratio or initial secondary mass distribution. We note, however, that
the theoretically predicted ratios are derived under the assumption of
a constant envelope binding-energy parameter and a constant
envelope-ejection parameter. In addition, the derivation of
constraints on the common-envelope ejection efficiency requires the
detection of WDMS binaries with luminosity ratios down to $10^{-5}$,
which poses a severe and not easily overcome observational challenge.

This paper is to serve as a first comprehensive step in a study of
more specific subclasses of WDMS binaries and binaries descending from
them. In future investigations we will address WDMS binaries in the
context of pre-cataclysmic variables, double white dwarfs (Willems et
al., in preparation), and white dwarf + B-star binaries (Willems et
al., in preparation). Other potentially interesting applications are
the study of type Ia supernovae progenitors through both the
'standard' evolutionary channel and the recently proposed thermal time
scale mass-transfer channel (King et al. 2003).

\section*{Acknowledgements}
We are grateful to Jarrod Hurley, Onno Pols, and Chris Tout for
sharing their SSE software package and to the referee, Gijs Nelemans,
whose constructive report and valuable suggestions contributed to the
improvement of the paper. Rob Mundin contributed to the subroutine
that calculates the common envelope evolution. This research was
supported by the British Particle Physics and Astronomy Research
Council (PPARC) and made use of NASA's Astrophysics Data System
Bibliographic Services.

\end{document}